\pdfoutput=1
\documentclass[a4paper,fleqn]{cas-dc}

\usepackage[numbers]{natbib}
\usepackage{soul}
\usepackage{siunitx}
\usepackage{hyperref}
\usepackage{adjustbox}

\usepackage{subcaption}

\def\tsc#1{\csdef{#1}{\textsc{\lowercase{#1}}\xspace}}
\tsc{WGM}
\tsc{QE}

\newcommand{\rebuttal}[1]{\textcolor{black}{#1}}
\newcommand{\srebuttal}[1]{\textcolor{black}{#1}}

\begin{document}
\let\WriteBookmarks\relax
\def\floatpagepagefraction{1}
\def\textpagefraction{.001}

\shorttitle{Morphing the left atrium geometry: The role of the pulmonary veins on flow patterns and thrombus formation}

\shortauthors{S. Rodríguez-Aparicio \textit{et al.}}  

\title [mode = title]{Morphing the left atrium geometry: The role of the pulmonary veins on flow patterns and thrombus formation}
%

\tnotemark[<tnote number>] 


%

\author[1]{Sergio Rodríguez-Aparicio}[orcid=0000-0001-7990-5969]

\affiliation[1]{organization={Departamento de Ingeniería Mecánica, Energética y de los Materiales, Universidad de Extremadura},
            addressline={Avda.de Elvas s/n}, 
            city={Badajoz},
            postcode={06006}, 
            country={Spain}}

\author[1,2]{ Conrado Ferrera}[orcid=0000-0002-2274-1374]

\affiliation[2]{organization={Instituto de Computación Científica Avanzada (ICCAEX)},
            addressline={Avda.de Elvas s/n}, 
            city={Badajoz},
            postcode={06006}, 
            country={Spain}}

\author[3]{ María Eugenia Fuentes-Cañamero}

\affiliation[3]{organization={Servicio de Cardiología, Hospital Universitario de Badajoz},
            addressline={Avda.de Elvas s/n}, 
            city={Badajoz},
            postcode={06006}, 
            country={Spain}}

\author[4]{ Javier García García}

\author[4]{ Jorge Dueñas-Pamplona}[orcid=0000-0002-6570-6624]
\cormark[1]
\ead{jorge.duenas.pamplona@upm.es}

\affiliation[4]{organization={Departamento de Ingeniería Energética, Universidad Politécnica de Madrid},
            addressline={Avda. de Ramiro de Maeztu 7}, 
            city={Madrid},
            postcode={28040}, 
            country={Spain}}

\cortext[4]{Corresponding author}

\begin{abstract}
\noindent\textbf{Background}: \rebuttal{Despite the significant advances made in the field of computational fluid dynamics (CFD)} to simulate the left atrium (LA) in atrial fibrillation (AF) conditions, the connection between atrial structure, flow dynamics, and blood stagnation in the left atrial appendage (LAA) remains unclear. Deepening our understanding of this \rebuttal{relationship} would have important clinical implications, as the thrombi formed within the LAA are one of the main causes of stroke.

\noindent\textbf{Aim}: \rebuttal{To highlight and better understand the fundamental role of the PV orientation in forming atrial flow patterns and systematically quantifying its effect on blood stasis within the LAA}. 

\noindent\textbf{Methods}: Two patients with \rebuttal{opposite} atrial flow patterns were selected for the study. The atria were segmented and subsequently morphed to modify the \rebuttal{pulmonary vein (PV) orientations} in a highly controlled manner. CFD analysis \rebuttal{were} performed using a kinematic model \rebuttal{able to reproduce AF conditions}. Results were projected into the universal left atrial appendage coordinate (ULAAC) system to enhance data visualization and comparison.

\noindent\textbf{Results}: \rebuttal{The position of the main atrial vortex can be modified by controlled changes in the PV orientations, which to the best of our knowledge was not demonstrated before. This finding may have important clinical implications, as the behavior and position of the main atrial vortex is crucial to define the LA flow patterns and thus the LAA washing, making possible to assess the stroke risk for a particular patient.}
\end{abstract}




\maketitle

\section{Introduction}\label{sec:intro}

Atrial fibrillation (AF) is the most common type of supraventricular arrhythmia \cite{SGC22}. In 2019, this disease affected 33.1 million people worldwide and caused 219.4 thousand deaths \cite{JLLWGM23}. Its prevalence has increased by 33\% in the past two decades, and this trend is expected to continue for the next three decades \cite{LSC21}.

AF appears as a result of an alteration in the electrical impulses of the pulmonary veins (PV). As a consequence, \rebuttal{the} atrial contraction becomes irregular \cite{SBKJH20}\rebuttal{. This usually occurs in conjunction with atrial remodeling, which affects} both the biomechanical and electrical functions of the myocardium \rebuttal{in the medium and long term}. These processes lead to increased blood stasis in the left atrial appendage (LAA) \cite{boyle2021fibrosis}.

The LAA is a lobe-like structure within the left atrium (LA), originated during embryonic atrial development \cite{Fukui_embryos24,Rajiah_CT_devices21}. Its morphology is highly patient-specific \cite{lupercio2016left}, \rebuttal{and works} as a contractile reservoir or as a decompression chamber depending on the phase of the cardiac cycle \cite{hoit2014left}. \rebuttal{This contraction is impaired in AF conditions, becoming the cause of thrombosis in up to 91\% of non-valvular AF patients \cite{Rajiah_CT_devices21}. Therefore,} patients with AF have a five-fold \rebuttal{higher} risk of stroke than those not affected by this disease \cite{Wolf1991,Rajiah_CT_devices21}.

The electrical isolation of PV and the closure of LAA are standard treatments for AF \cite{Gonzalez-Casal_anatomy_22, SGC22}. Closure of LAA is a common practice to prevent stroke in patients \rebuttal{with no positive anticoagulant response} \cite{Gonzalez-Casal_anatomy_22,Rajiah_CT_devices21}. However, there is a group of patients who have AF recurrence after electrical isolation \cite{D16} or device-related thrombus after LAA closure \cite{Fauchier2018,Aminian2019}. Therefore, a deeper understanding of \rebuttal{the} atrial flow patterns is required for \rebuttal{stratifying the patients and developing new treatments.}

Current imaging techniques, such as transesophageal echocardiography and 4D Magnetic Resonance Imaging, provide useful but limited flow information within the LAA \cite{valvez2023,mill2023role}. \rebuttal{This information could be obtained by solving the Navier-Stokes equations, which provide a 3D time-resolved flow representation. In this regard, the most common approach when solving cardiovascular flows is computational fluid dynamics (CFD). Although there are recent tools from fractional calculus very useful to provide feedback on certain diseases \cite{maayah2022numerical,maayah2022multistep}, flow resolution is still extremely challenging \cite{sharma2024variational}. CFD allows exploring flow patterns in complex geometries such as the heart, ranging from one-cavity models} \cite{chnafa2014image, Otani2016,Lantz2018a},  to whole left heart models \cite{Vedula2015,bucelli2022mathematical, mihalef2011patient}. \rebuttal{CFD models} can even integrate the fluid-structure interaction with electrophysiological or biomechanical mechanisms \cite{bucelli2022mathematical, corti2022impact, gonzalo2022non, feng2019analysis}.

\rebuttal{Regarding AF,} many \rebuttal{CFD} studies have focused on the analysis of LAA stasis \cite{Otani2016,Bosi2018,Garcia-Isla2018,Masci2019,duenas2021comprehensive,Koizumi2015,corti2022impact,garcia2021demonstration,duran2023pulmonary}. \rebuttal{As there} is clinical evidence on the relationship between LAA morphology and stroke risk in patients with AF \cite{DiBiase2012, yamamoto2014complex}\rebuttal{, some of these studies have analyzed} the link between LAA stasis and LAA morphology \cite{Masci2019, Garcia-Isla2018, garcia2021demonstration, duenas2022morphing}. However, \rebuttal{the subjective component associated with the radiologist limits the reproducibility of the proposed morphological classifications} \cite{Wang2010,yaghi2020left}.

\rebuttal{Other} researchers have studied specific variables within LAA such as volume, width, height, bending angles, tortuosity, ostium diameters, etc. \cite{Jeong2016,pons2022joint,beinart2011left,lee2017additional,yamamoto2014complex,Mill2021}. \rebuttal{Reduced-order models have also been proposed} to extract the specific flow features \cite{duenas2024reduced}.

 \rebuttal{Moreover, other factors such as cardiac condition, remodeling processes, and rheology \cite{boyle2021fibrosis} are involved} in the thrombogenic process within LAA. Due to its \rebuttal{complex} and multifactorial nature, this problem has been addressed by \rebuttal{assuming different hypothesis:} (a) a rigid LA wall \cite{duenas2021comprehensive,garcia2021demonstration}, (b) blood as a Newtonian fluid \cite{gonzalo2022non}, (c) different boundary conditions \cite{duenas2021boundary} or (d) different flow-split ratios \cite{duran2023pulmonary,rodriguez2024influence}.

\rebuttal{Although the mentioned studies cover a significant number of factors known to influence the atrial flow, little attention has been paid to the influence of PV orientations. To the best of our knowledge, no study has analyzed the isolated effect of varying the PV orientation on atrial flow patterns. We aim to assess the influence of this orientation on the formation and position of the main atrial vortex, and specifically to assess how it is related to stasis within the LAA. In this paper, we applied some of the most recent techniques in the field to perform this analysis: morphing atrial geometries to gradually modify the PV orientations, employing a kinematic model to recreate the fibrillated atrial motion \cite{zingaro2021geometric,zingaro2021hemodynamics, corti2022impact}, and projecting the results on the ULAAC system \cite{duenas2024reduced} to allow comparison between different morphologies. The main contributions of this paper are:}
\begin{enumerate}
\item{\rebuttal{Morph two atrial geometries to gradually modify the PV orientations, performing the transition between two extreme atrial flow patterns.}}
\item{\rebuttal{Establish the relationship between PV orientation, atrial flow patterns and stasis in the LAA.}}
\item{\rebuttal{Study how changing the PV orientations affects hemodynamic indices and vortical structures.}}

\end{enumerate}

\section{\rebuttal{Related work}}

\rebuttal{The orientation of the PV is clinically a critical factor because it hinders catheter manipulation during the ablation procedure and affects AF recurrence after the procedure \cite{Szegedi21,Gal2015,Gal2017}.} From an anatomical point of view, \rebuttal{although the number of PV ranges between 3 and 7 \cite{Fukui_embryos24,Mill2021_pulmonary,Marom2004}, around 70 \% of the population has 4 PV \cite{SGC22,Rajiah_CT_devices21,Marom2004}. Moreover, the PV location} within the \rebuttal{atrial cavity is known to affect} the flow and the velocity profile of the MV \cite{dahl2012,Fang2021, rodriguez2024influence}. In fact, a comparison between the groups with/without a history of stroke has shown a difference in the angle between the orifices of the left/right PV and the LAA ostium in both groups \cite{Fang_2022}. Therefore, the orientation of the PV is a \rebuttal{relevant parameter when assessing the stroke risk}.

However, the number of CFD studies that include it as a parameter to assess stroke risk is reduced. \rebuttal{Dueñas-Pamplona} et al. \cite{duenas2022morphing} morphed an atrium to \rebuttal{obtain the mean and extreme physiological PV orientations} previously reported in the bibliography. The results confirmed how variations in the orientation of the PV can affect stasis within the LAA. Recent studies \cite{Mill2021_pulmonary, mill2023role}, with a large cohort of patients, reported that the angle between left PV (LPV) and the angle between RPV influence the formation of flow structures within the LA. These studies also pointed out that the orientation of the RPV determines if the atrial flow is oriented directly to the MV, as also shown in \cite{rodriguez2024influence}. All of these studies \cite{Mill2021_pulmonary, duenas2022morphing,mill2023role, rodriguez2024influence} confirmed that the influence of the angles of PV on blood stasis should not be ignored under AF conditions, highlighting the need to better understand the relationship between PV orientations and blood stasis within the LAA. \rebuttal{Nevertheless, these studies \cite{Mill2021_pulmonary,mill2023role} combined} several structural modifications (LAA and LA morphologies, PV locations, and orientations), so the isolated effect of PV orientations was not analyzed.

To our knowledge, only one study \rebuttal{has investigated} the isolated \rebuttal{effect} of varying the orientation of the PV \cite{duenas2022morphing}. \rebuttal{This was achieved applying a morphing technique to some patient-specific geometries to introduced highly controlled anatomical modifications. However, the work did not focus on understanding how flow patterns were modified, so the influence of the PV orientation was not studied in depth.} A recent study has observed how different representative atrial flow patterns can be identified depending on the position of the main atrial vortex within the LA \cite{rodriguez2024influence}. They observed how the main atrial vortex varies its position with the flow split ratio, suggesting that its position may also vary depending on the PV orientation.

\rebuttal{In this paper, two patients were selected from the previous literature with the most different atrial flow patterns \cite{rodriguez2024influence}. The morphing technique was applied to gradually modify the PV orientations. The transitions were performed in a parametric manner between both flow patterns,} demonstrating how the PV orientation plays a fundamental role in the formation of atrial flow patterns. \rebuttal{And even more}, showing in which way the orientations of LPV and RPV affect LAA stasis, and thus may influence the risk of thrombosis in the medium and long term.

\section{Methods}
\label{sec:methods}
\begin{figure*}[t]
\centerline{\includegraphics[width=0.8\linewidth]{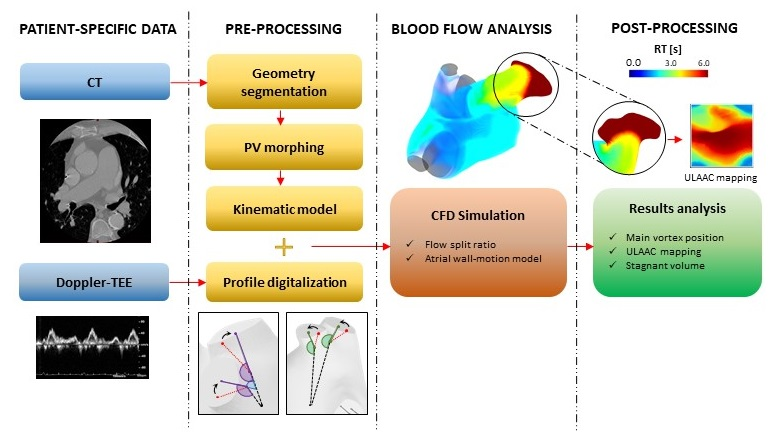}}
\caption{Workflow of the study. \textit{Patient-specific data:} The initial data were CT images and Doppler TEE. \textit{Data pre-processing:} CT images were segmented. The resulting atrial models were morphed to obtain several versions with different PV orientations. Boundary conditions were obtained for each case from a \rebuttal{kinematic} model, a CFD-ready volumetric mesh, and PV Doppler curves. \textit{Flow analysis:} Two simulations with different flow split ratios were run for each case. \textit{Post-processing:}  The ULAAC system was used to \rebuttal{visualize and compare results between the different cases}.} 
\label{fig:workflow}
\end{figure*}

As mentioned, this work aims to study the influence of the orientation of PV on LA flow, as it can strongly influence blood renewal within LAA. \rebuttal{In a previous work \cite{rodriguez2024influence}, it was found that a cohort of nine atrial geometries can be classified into two representative atrial flow patterns depending on the stasis response to changes in the flow-split ratio: types A and B. Interestingly, this response was reported to depend on the position of the center of the main circulatory flow: When it is located right after the LPV, the atrial flow pattern is Type A, whereas if it is located near the LAA ostium or close to the LA roof, it is Type B. To demonstrate how the PV orientations define the formation of the atrial flow patterns, 
we selected two representative patients with opposite flow patterns (A and B), and modified the PV angles of a Type A atrium to obtain a Type B behavior (transition AB) and vice versa (transition BA).} The complete framework involved CT scanning, geometry segmentation, anatomical morphing, Doppler TEE digitalization, wall motion implementation, CFD simulation, and post-processing. The current workflow is summarized in Figure \ref{fig:workflow}.

\subsection{Clinical data}
\label{subsec:medical_data}

As mentioned, two patients with \rebuttal{opposite} atrial flow patterns were selected from our database (Table \ref{tab:clinic-data}). Both participants provided their informed written consent according to the guidelines of the Declaration of Helsinki. The Ethics Committee of the University Hospital of Badajoz (Spain) and the University of Extremadura authorized the study.

\begin{table*}[b]
\centering
\caption{\rebuttal{Clinical data}}

\begin{tabular}{|l|ccccccc|}
\hline

  \multicolumn{1}{|l|}{}  & \textbf{Age} & \textbf{Sex} & \textbf{Clinic history} & \textbf{Af type} & \textbf{LA volume} & \textbf{HR} & \textbf{LAA volume} \\ 

    \multicolumn{1}{|l|}{}  & \textbf{[years]} & \textbf{[M/F]} & \textbf{} & \textbf{} & \textbf{[mL]} & \textbf{[bpm]} & \textbf{[mL]} \\ \hline
 
\textbf{Patient 1 - Transition AB} & 89 & F & HT/DLP & PE & 114-138 & 50 & 10 \\ \hline
\textbf{Patient 2 - Transition BA} & 83 & M & HT & PA & 153-175 & 75 & 19 \\ \hline

\end{tabular}
\\ \rebuttal{HT - hypertension; DLP - cholesterol;  HR - heart rate; PE - Permanent; PA - Paroxysmal}
\label{tab:clinic-data}
\end{table*}

The velocity in each PV and mitral valve (MV) was measured using EPIC CVx and Affiniti CVx cardiovascular ultrasound systems (Philips, The Netherlands). An in-house code, written in Matlab $\text{\textregistered}$ (Mathworks, Inc., Natick, MA, USA), digitalized these velocity profiles.

CT scans were obtained using a LightSpeed VCT General Electric Medical Systems scanner (Milwaukee, WI, USA). The Doppler waveforms were synchronized as the TEE and CT scans \rebuttal{were not acquired simultaneously}. PV measurements were used to impose boundary conditions for the inflow, while MV measurements were used to validate the simulation.

\begin{figure*}[t]
\centering
   \begin{subfigure}[t]{0.35\textwidth}
        \centering
        \includegraphics[width=\linewidth]{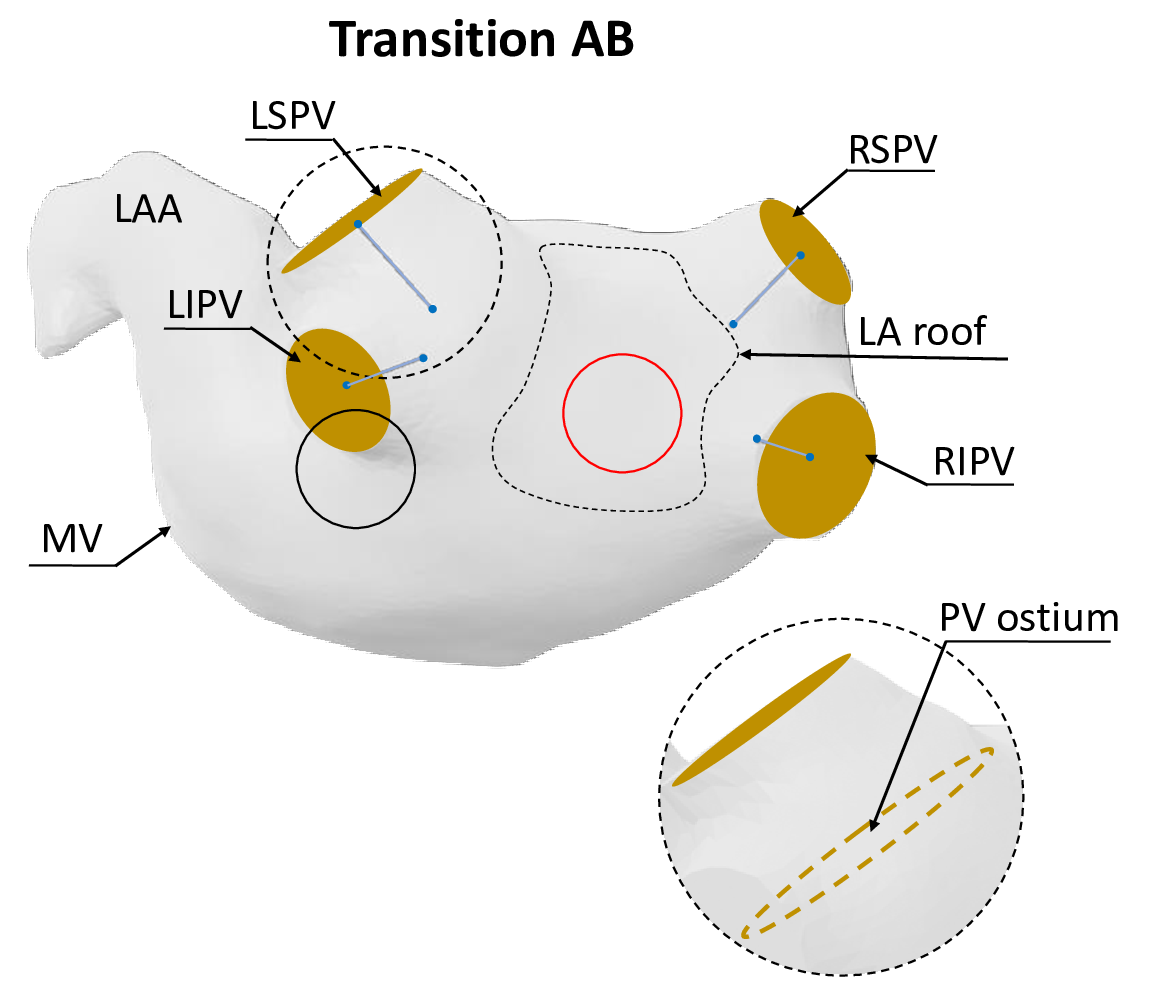}
        \caption{}
          \label{subfig:presentation_atb}
    \end{subfigure}
       \begin{subfigure}[t]{0.35\textwidth}
        \centering
        \includegraphics[width=\linewidth]{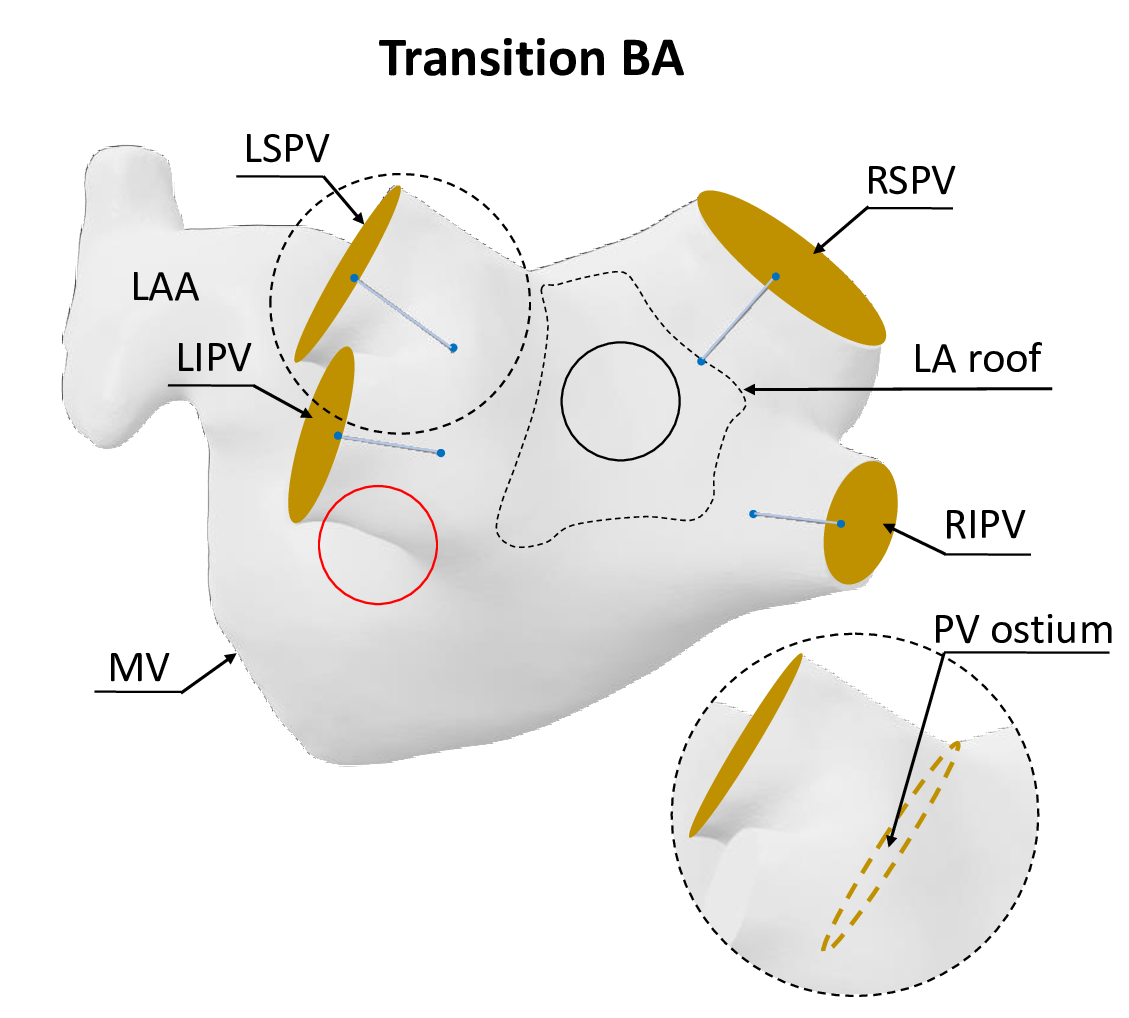}
        \caption{}
          \label{subfig:presentation_bta}
    \end{subfigure}
\caption{Patient-specific geometries for both patients. Regions enclosed by dashed lines represent LA roof areas. Black and red circles represent the initial and final position of the main circulatory flow, respectively. A yellow dashed circle depicts the PV ostium area in the zoomed region.}
\label{fig:cases_parts}
\end{figure*}

Each set of CT images was segmented following a semi-automatic procedure \cite{duenas2021comprehensive}. The segmentation process was supervised by two experts, a radiologist and an interventionist cardiologist. Two surface meshes were generated, one for each anatomy in 0\%RR. RR is the time between two consecutive electrocardiographic R waves, i.e. patient-specific duration of the cardiac cycle, where 0\%RR stands for electrocardiographic end-diastole of the ventricle. An open-source 3D software, Blender 3.2 \cite{Blender2023}, was used to clean and adjust the meshes for a subsequent simulation. Figure \ref{fig:cases_parts} shows the geometry \rebuttal{that results} for each patient.

\begin{figure*}[t]
\centering
   \begin{subfigure}[t]{0.35\textwidth}
        \centering
        \includegraphics[width=\linewidth]{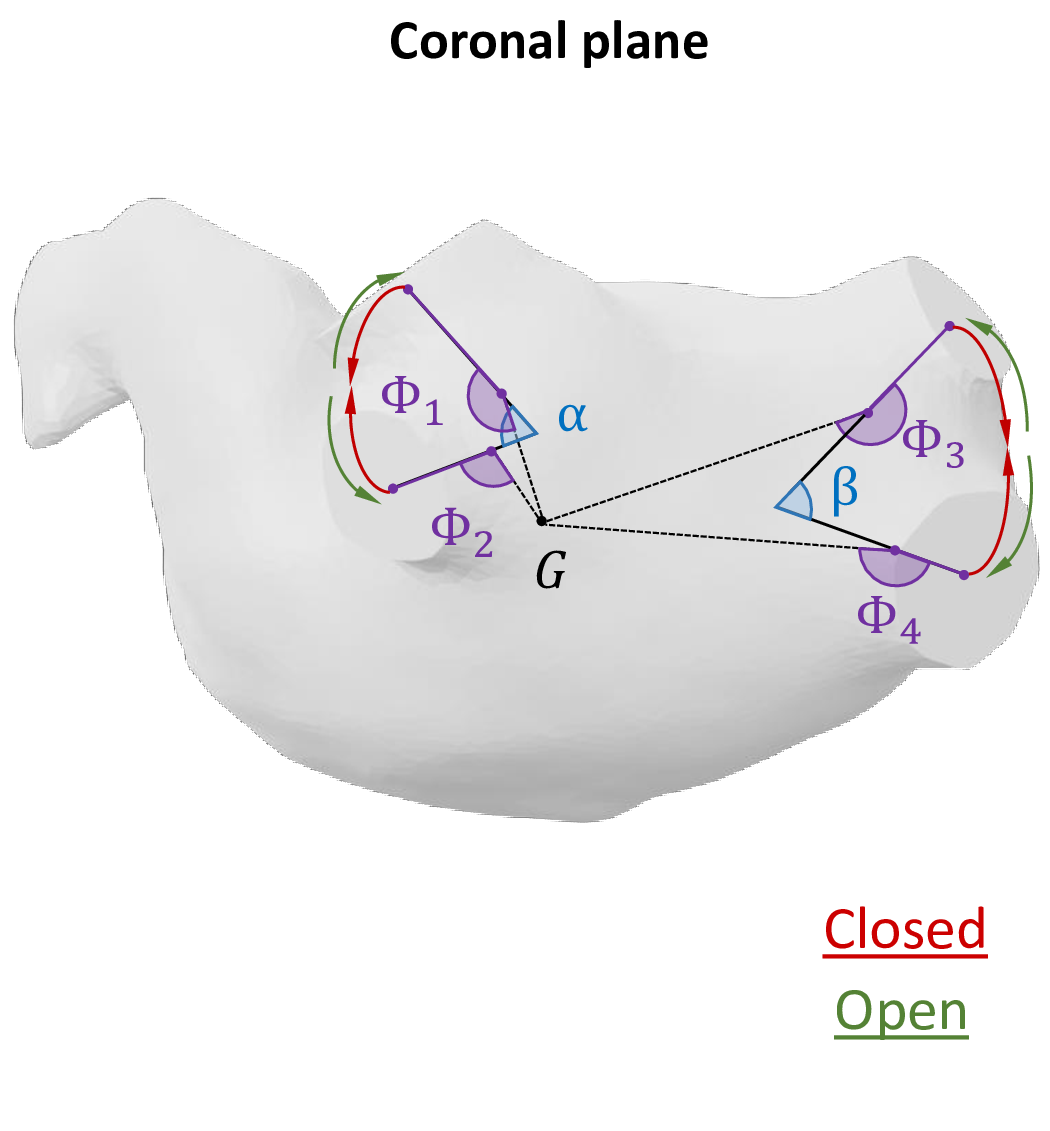}
        \caption{}
          \label{subfig:angles_coronal}
    \end{subfigure}
       \begin{subfigure}[t]{0.35\textwidth}
        \centering
        \includegraphics[width=\linewidth]{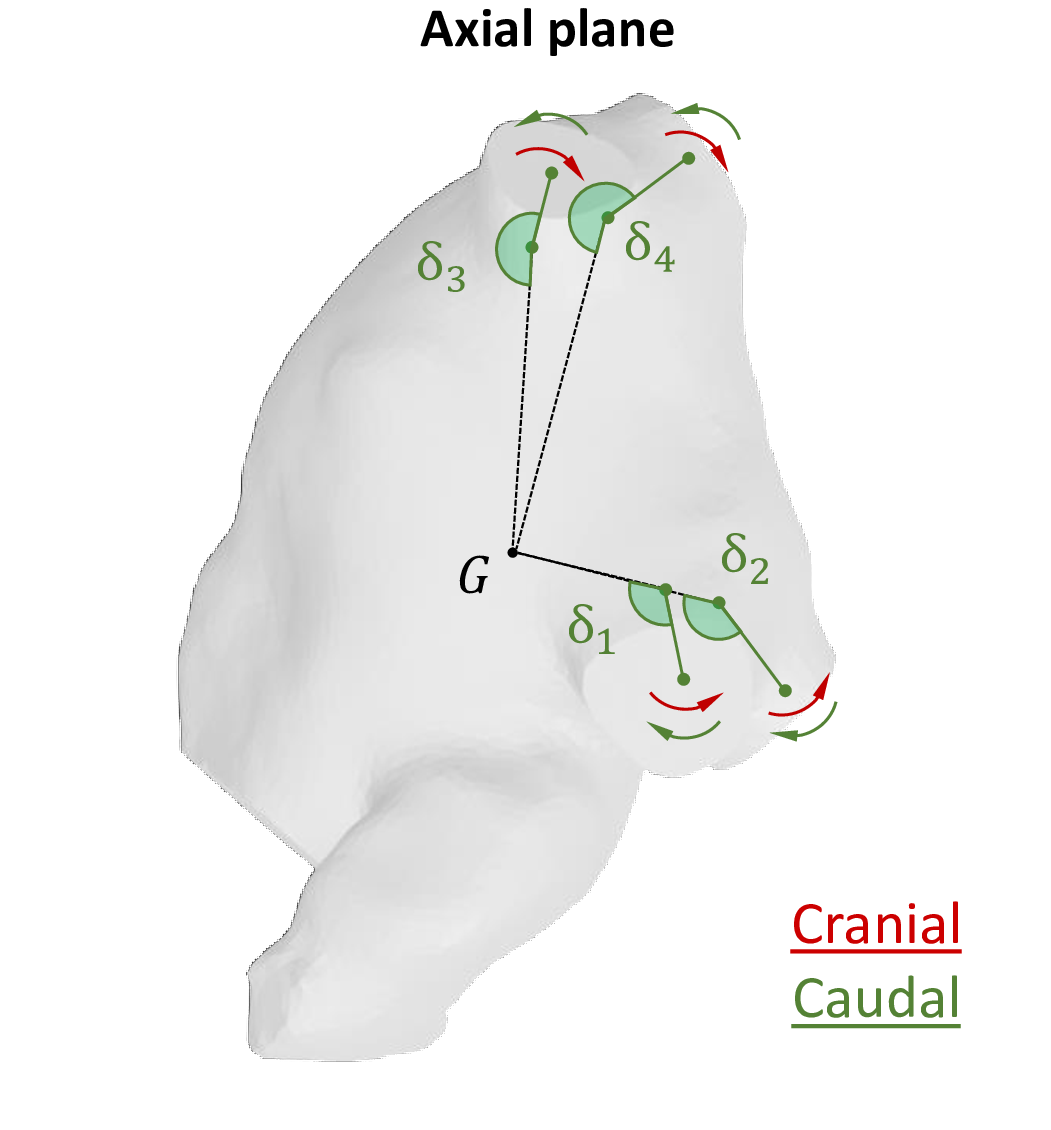}
        \caption{}
          \label{subfig:angles_axial}
    \end{subfigure}

\caption{(a) \textbf{$\Phi_i$}, $\boldsymbol{\alpha}$, and $\boldsymbol{\beta}$ angles defined in the coronal plane. (b) $\boldsymbol{\delta_i}$ angle defined in the axial plane.}
\label{fig:angles}
\end{figure*}

\begin{figure*}[t]
 \centering
   \begin{subfigure}[t]{0.43\textwidth}
        \centering
        \includegraphics[width=\linewidth]{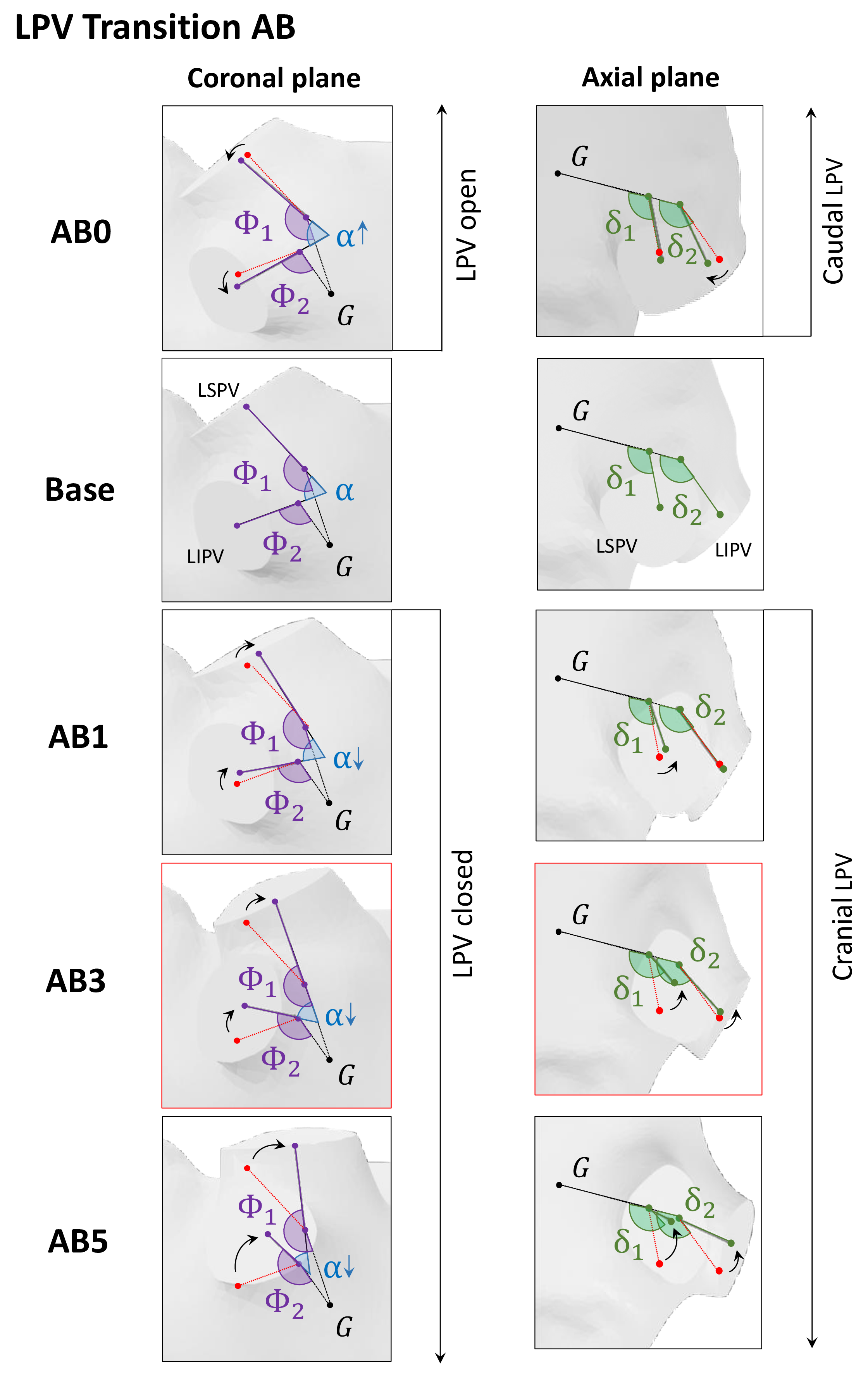}
        \caption{}
          \label{subfig:cases_atb1}
    \end{subfigure}
       \begin{subfigure}[t]{0.43\textwidth}
        \centering
        \includegraphics[width=\linewidth]{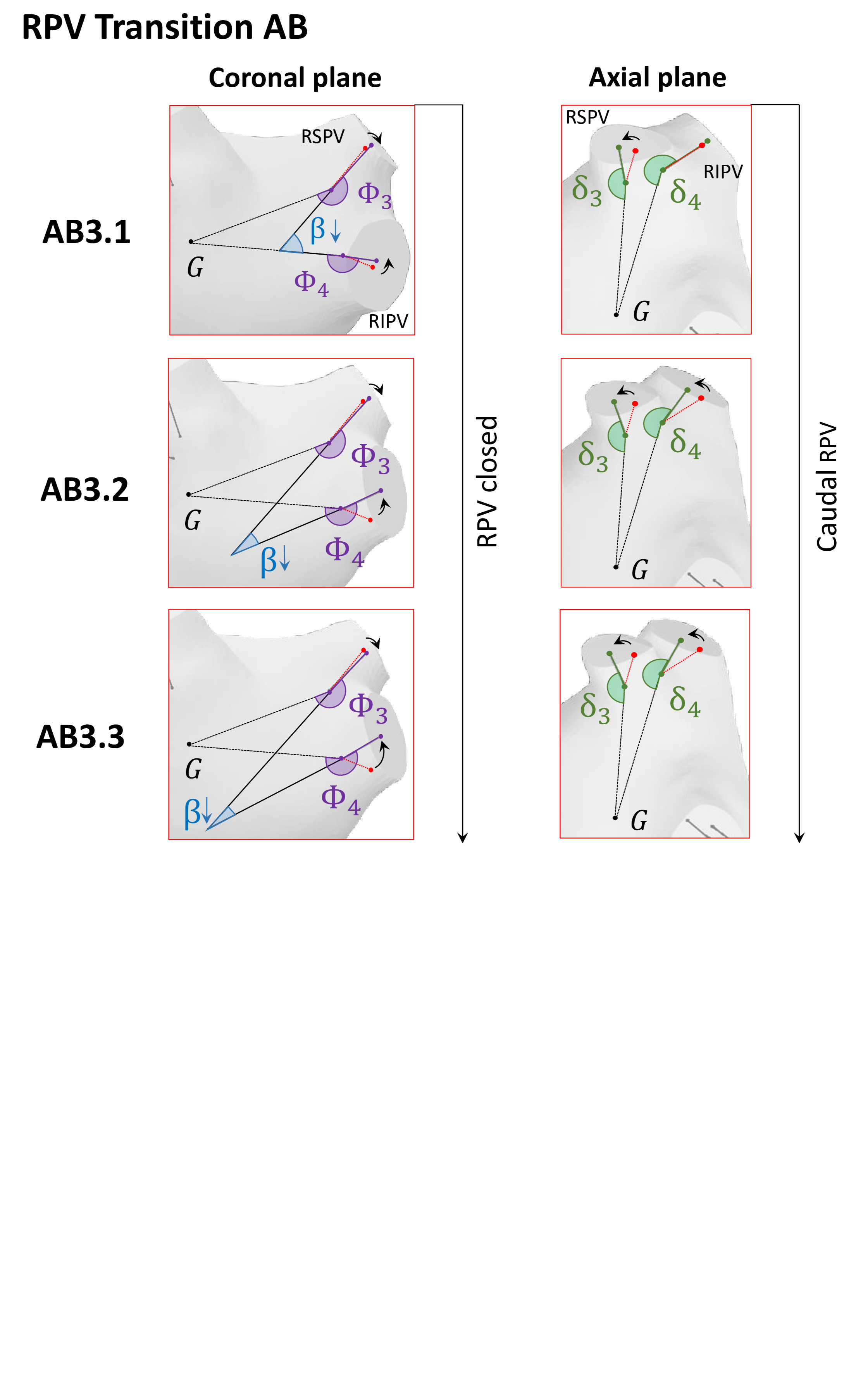}
        \caption{}
          \label{subfig:casesatb2}
    \end{subfigure}
    
\caption{PV angle transformations in the coronal (purple) and axial plane (green) with respect to baseline case segments (in red) for Transition AB. (a) LPV transformations to change the position of the main atrial circulatory flow, and (b) RPV transformations to direct the RPV flow to the new position of the main atrial circulatory flow. \rebuttal{A supplementary figure is included (Supplementary Data A.1.1).}} 

\label{fig:geoms_atb}
\end{figure*}

\begin{table*}[b]
\centering
\caption{\rebuttal{Angular combinations applied for transitioning a type A case flow pattern into a type B.}}

\begin{tabular}{|l|cccccc|cccc|}
\hline
\multicolumn{11}{|c|}{Transition AB} \\ \hline
\multicolumn{1}{|l|}{\multirow{2}{*}{Cases}} & \multicolumn{6}{c|}{Coronal plane}  & \multicolumn{4}{c|}{Axial plane} \\ 
 & \textbf{$\Phi_1$} & \textbf{$\Phi_2$} & \textbf{$\Phi_3$} & \textbf{$\Phi_4$} & $\boldsymbol{\alpha}$ & $\boldsymbol{\beta}$ &  $\boldsymbol{\delta_1}$ & $\boldsymbol{\delta_2}$ & $\boldsymbol{\delta_3}$ & $\boldsymbol{\delta_4}$ \\ \hline
 
\textbf{AB0} & 140 & 90 & 211 & 170 & 72 & 65 & 113 & 132 & 192 & 219\\ \hline
\textbf{Base} & 155 & 105 & 211 & 170 & 70 & 65 & 118 & 140 & 192 & 219 \\ \hline
\textbf{AB1} & 167 & 118 & 211 & 170 & 66 & 65 & 123 & 148 & 192 & 219 \\ \hline
\textbf{AB2} & 175 & 124 & 211 & 170 & 66 & 65 & 132 & 150 & 192 & 219\\ \hline
\textbf{AB3} & 177 & 132 & 211 & 170 & 62 & 65 & 140 & 152 & 192 & 219 \\ \hline
\textbf{AB4} & 183 & 144 & 211 & 170 & 57 & 65 & 148 & 158 & 192 & 219\\ \hline
\textbf{AB5} & 191 & 167 & 211 & 170 & 43 & 65 & 160 & 169 & 192 & 219 \\ \hline\hline
\textbf{AB3.1} & 177 &132 & 207 & 175 & 62 & 54 & 140 & 152 & 167 & 218 \\ \hline
\textbf{AB3.2} & 177 & 132 & 205 & 207 & 62 & 21 & 140 & 152 & 160 & 197 \\ \hline
\textbf{AB3.3} & 177 & 132 & 205 & 214 & 62 & 14 & 140 & 152 & 155 & 189\\ \hline

\end{tabular}
\label{tab:angle-data1}
\end{table*}

\begin{figure*}[t]
 \centering
   \begin{subfigure}[t]{0.43\textwidth}
        \includegraphics[width=\linewidth]{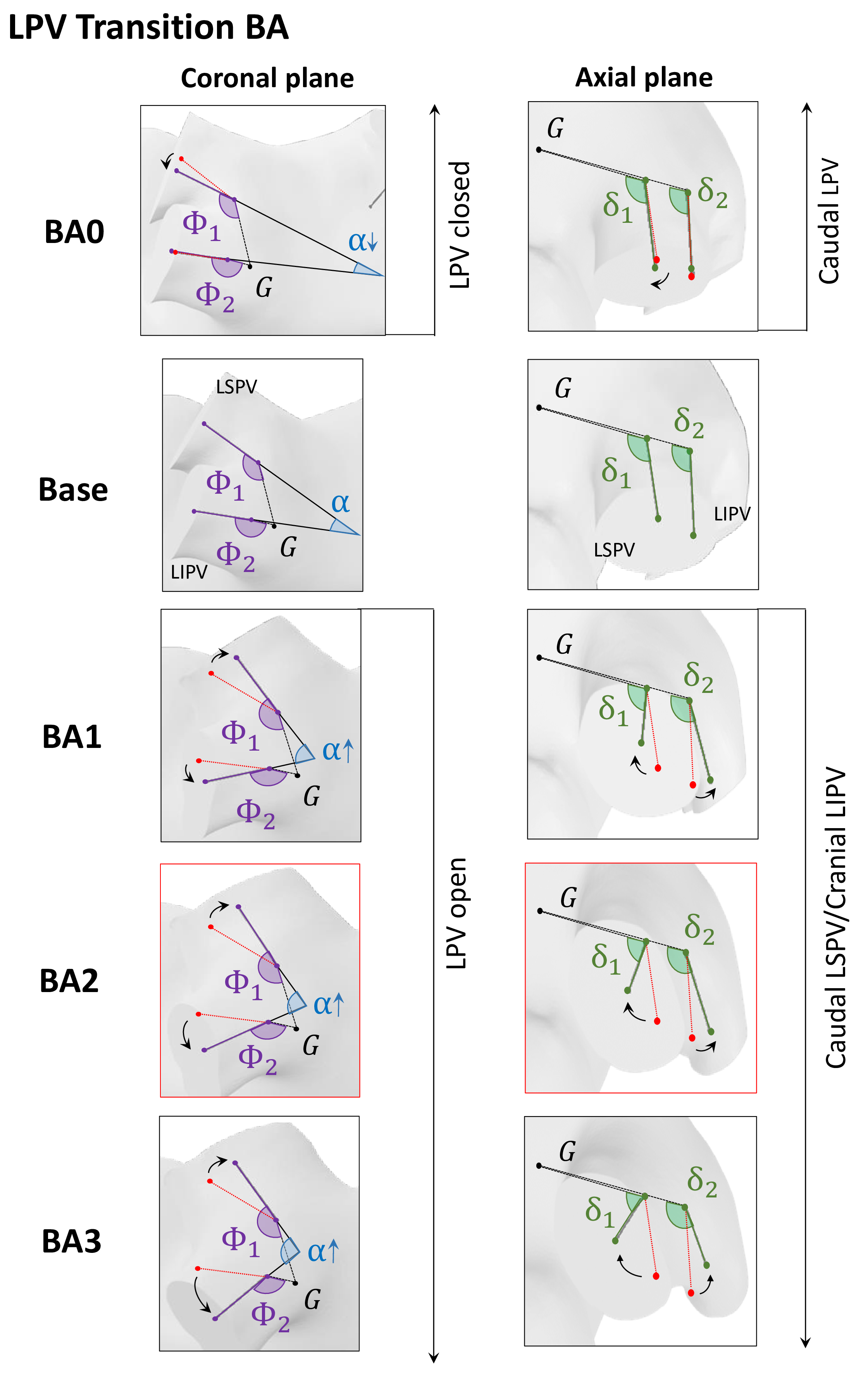}
        \caption{}
          \label{subfig:cases_bta1}
    \end{subfigure}
       \begin{subfigure}[t]{0.43\textwidth}
        \centering
        \includegraphics[width=\linewidth]{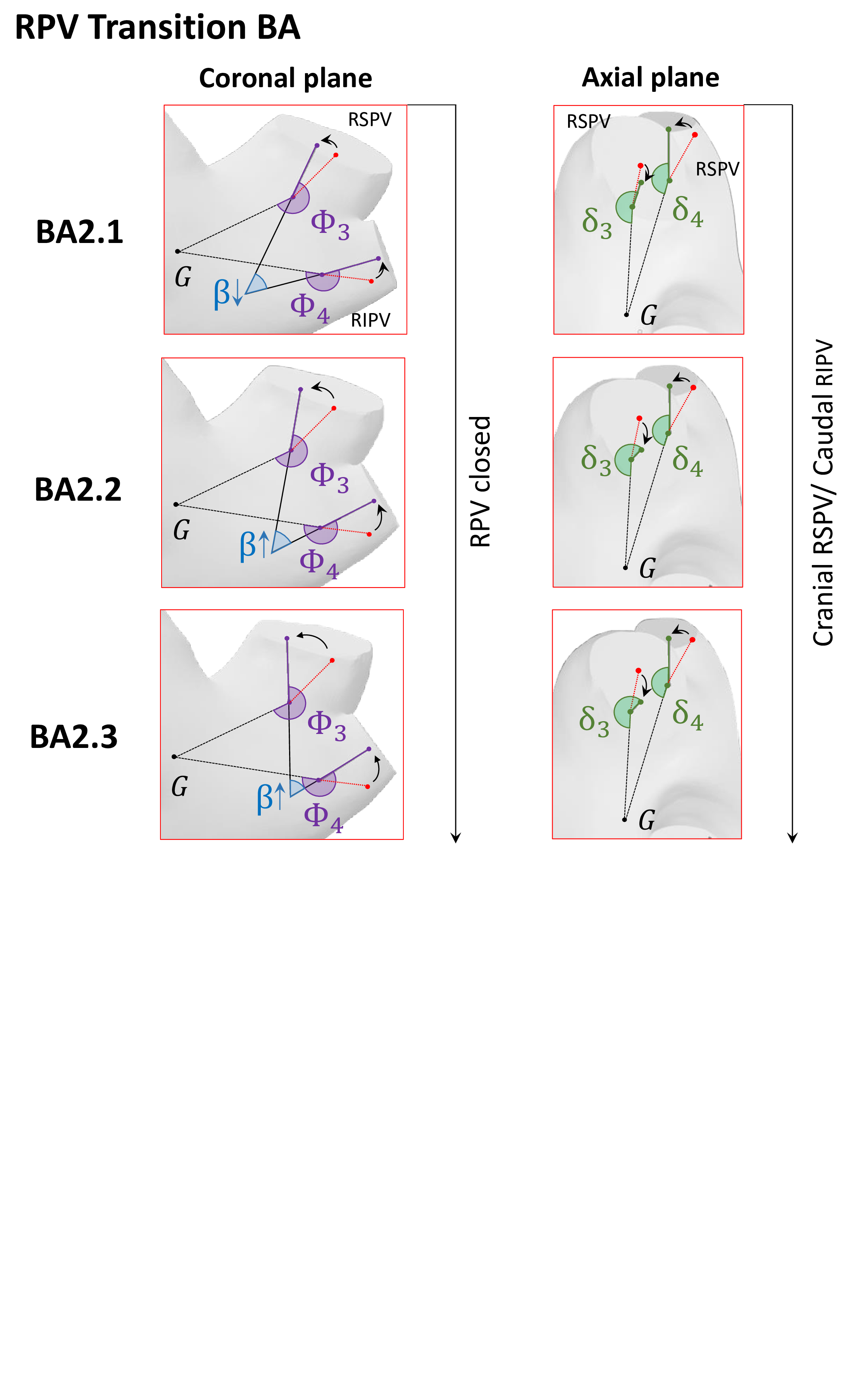}
        \caption{}
          \label{subfig:casesbta2}
    \end{subfigure}
    
\caption{PV angle transformations in the coronal (purple) and axial plane (green) with respect to baseline case segments (in red) for Transition BA. (a) LPV transformations to change the position of the main atrial circulatory flow, and (b) RPV transformations to direct the RPV flow to the new position of the main atrial circulatory flow. \rebuttal{A supplementary figure is included (Supplementary Data A.1.2).}}
\label{fig:geoms_bta}
\end{figure*}

\begin{table*}[b]
\centering
\caption{Angular combinations applied for transitioning a type B case flow pattern into a type A.}

\begin{tabular}{|l|cccccc|cccc|}
\hline
\multicolumn{11}{|c|}{Transition BA} \\ \hline
\multicolumn{1}{|l|}{\multirow{2}{*}{Cases}} & \multicolumn{6}{c|}{Coronal plane}  & \multicolumn{4}{c|}{Axial plane} \\ 
  \multicolumn{1}{|l|}{}  & \textbf{$\Phi_1$} & \textbf{$\Phi_2$} & \textbf{$\Phi_3$} & \textbf{$\Phi_4$} & $\boldsymbol{\alpha}$ & $\boldsymbol{\beta}$ &  $\boldsymbol{\delta_1}$ & $\boldsymbol{\delta_2}$ & $\boldsymbol{\delta_3}$ & $\boldsymbol{\delta_4}$ \\ \hline
 
\textbf{BA0} & 131 & 168 & 198 & 179 & 16 & 54 & 120 & 99 & 175 & 139\\ \hline
\textbf{Base} & 144 & 166 & 198 & 179 & 27 & 54 & 108 & 101 & 175 & 191\\ \hline
\textbf{BA1} & 164 & 146 & 198 & 179 & 71 & 54 & 94 & 118 & 175 & 191\\ \hline
\textbf{BA2} & 173 & 129 & 198 & 179 & 95 & 54 & 81 & 128 & 175 & 191\\ \hline
\textbf{BA3} & 173 & 112 & 198 & 179 & 111 & 54 & 63 & 128 & 175 & 191\\ 
\hline\hline
\textbf{BA2.1} & 173 & 129 & 219 & 208 & 95 & 48 & 81 & 128 & 205 & 163\\ \hline
\textbf{BA2.2} & 173 & 129 & 236 & 216 & 95 & 54 & 81 & 128 & 220 & 162\\ \hline
\textbf{BA2.3} & 173 & 129 & 246 & 225 & 95 & 59 & 81 & 128 & 285 & 158\\ \hline

\end{tabular}
\label{tab:angle-data2}
\end{table*}

\subsection{Anatomical morphing}
\label{subsec:morphological_descriptors}
\subsubsection{Geometrical parametrization}

We used a methodology similar to that proposed by Mill \textit{et al.} \cite{mill2023role} to \rebuttal{define the characteristic angles of the PV}, but selecting as a reference the coronal and axial planes (Figure \ref{fig:angles}). We adopted \rebuttal{as a reference} these anatomical planes because they are widely used in common clinical practice\rebuttal{, facilitating the comparison between} different anatomies \cite{duenas2022morphing}. The angles were defined as follows. Two segments were defined for each PV and reference plane, one segment from the center of the PV cut (yellow surface in Figure \ref{fig:cases_parts}) to the center of the PV ostium (yellow dashed lines in Figure \ref{fig:cases_parts}), and another segment from this point to the atrial center of gravity. Angles $\Phi_i$ and $\delta_i$ were defined between the PV segments in the coronal plane (Figure \ref{subfig:angles_coronal}) and the axial plane (Figure \ref{subfig:angles_axial}), respectively. Angles $\alpha$ and $\beta$ were defined in the coronal plane between LPV and RPV, respectively (Figure \ref{subfig:angles_coronal}). Figure \ref{fig:angles}
also shows the standard clinical criteria for referring to the different PV orientations: open/closed (coronal plane) and cranial/caudal (axial plane). 
\rebuttal{Finally, it is important to mention that $\alpha$ range is reported in the bibliography (63$^\circ$-91$^\circ$) \cite{mill2023role}. This range has been confirmed (58$^\circ$-86$^\circ$) in a recent study \cite{rodriguez2024influence}. Interestingly, Rodriguez et al. \cite{rodriguez2024influence} reported on an outlier whose $\alpha$ was 27$^\circ$. This case represents an extreme type B atrial flow pattern. That geometry and a usual one ($\alpha=70^\circ$), with type A flow pattern, are morphed in this study.}

\subsubsection{LA rigging and skinning}

The 3D segmented geometries were rigged and skinned following the method proposed by \rebuttal{Dueñas-Pamplona} \textit{et al.} \cite{duenas2022morphing} to obtain different PV orientations. We used Blender 3.2, an open-source 3D suite with full rigging and morphing capabilities \cite{Blender2023}. First, the rigging process provided each atrial geometry with an armature formed by different bones. For each atrial geometry, the following bones were defined: right superior PV (RSPV), right inferior PV (RIPV), left superior PV (LSPV), and left inferior
PV (LIPV). Each bone was defined between the center of the PV cut and the center of the PV ostium, shown in Figure \ref{fig:cases_parts} as blue segments. Second, for the skinning process, the \textit{envelope technique} was applied \cite{Blender2023}. This technique uses the \textit{heat algorithm} \cite{crane2017heat} to calculate the influence region for each bone based on its distance from the surrounding points. Each bone controlled the mesh vertices under its influence, allowing us to modify the PV orientations by selecting the characteristic angles $\Phi_i$ and $\delta_i$. 

\subsubsection{LA morphing}

 We will now describe the morphing process that has allowed us to transform the atrial flow patterns by only modifying the characteristic angles of PV: \textit{Transition AB}, from type A to type B, and \textit{ Transition BA}, from type B to type A.

\textit{Transition AB}: To perform this transition, the following two conditions must be met: (i) the main circulatory flow should be displaced from near the exits of the LPVs (black circle in Figure \ref{subfig:presentation_atb}) to the region near the LA roof and the RPVs (red circle in Figure \ref{subfig:presentation_atb}), and (ii) the RPVs should be oriented toward this new position of the main circulatory flow. Once the transformation is complete, it should be checked that the RT increases in the modified geometry when the flow split changes from 59-41\% to 37-63\% LPV-RPV, as this is the main feature of type B anatomies \cite{rodriguez2024influence}. \rebuttal{These ratios where chosen as the extreme values reported in the literature when the body is lying in the left (59-41\%) and right side (37-63\%) respectively \cite{WRAPU19}.}

Requirement (i) is gradually achieved by decreasing $\alpha$ and increasing $\Phi_1$ and $\Phi_2$, as can be seen in Figure \ref{subfig:cases_atb1} (cases AB1-5). Decreasing $\alpha$ reduces interference between LPV flows and, with increasing $\Phi_1$ and $\Phi_2$, allows the main circulatory flow to move toward the LA roof, far from the exits of the LPV.

Requirement (ii) is achieved by directing the RPV flow to the new position of the main circulatory flow, i.e., increasing $\Phi_4$ and decreasing $\Phi_3$, $\beta$, $\delta_3$, and $\delta_4$. Modifications to the characteristic angles of RPV for the transition AB are shown in Figure \ref{subfig:casesatb2} (cases AB3.1-3.3). Table \ref{tab:angle-data1} summarizes the characteristic angles of the morphed geometries generated during the AB transition.

\textit{Transition BA}: To perform this transition, the following
two conditions must be met: (i) the main circulatory flow should be displaced from the LA roof (black circle in Figure \ref{subfig:presentation_bta} to the region near the exits of the LPV (red circle in Figure \ref{subfig:presentation_bta}, and (ii) the RPV flow should be oriented toward the MV. Once the transformation is complete, it should be checked that the RT decreases in the modified geometry when the flow split changes from 59-41\% to 37-63\% LPV-RPV, as this is the main feature of type A anatomies \cite{rodriguez2024influence}. This is due to a slight displacement of the center of the circulatory flow toward the LAA ostium, which favors LAA blood renewal.

Requirement (i) is achieved by gradually decreasing $\Phi_2$ and increasing $\Phi_1$ and $\alpha$, as can be seen in Figure \ref{subfig:cases_bta1} (cases BA1-3). Increasing $\alpha$ brings forward the collision between the LPV flows and, together with decreasing $\Phi_2$ and increasing $\Phi_1$, allows the main circulatory flow to move towards the exits of the LPV, far from the LAA roof. Requirement (ii) is achieved by directing the RPV flow to the MV, increasing $\Phi_3$, $\Phi_4$, and $\delta_3$, and decreasing $\delta_4$. Modifications to the characteristic angles of the RPV for the BA transition are shown in Figure \ref{subfig:casesbta2} (cases BA2.1-2.3). 
Table \ref{tab:angle-data2} summarizes the characteristic angles of the morphed geometries generated during the BA transition.

\subsection{Parametrization of the LA wall motion}
\label{subsec:wall_motion}
Applying the transitions to patient-specific geometries results in a set of morphed geometries (10 for \rebuttal{Patient 1} and 8 for \rebuttal{Patient 2}). These generated geometries lack any information regarding the LA displacement field, so an \rebuttal{kinematic} model was employed to recreate the atrial wall motion. This displacement was generated following the methodology described by Zíngaro \textit{et al.} \cite{zingaro2021hemodynamics}, which was also employed in several recent works \cite{corti2022impact,rodriguez2024influence, zingaro2021geometric, kjeldsberg2024impact}. Following this method, we calculated the atrial boundary velocities \rebuttal{that generate the volume variation deduced from} each patient-specific flow profile.

\subsection{\rebuttal{Numerical model}}
\label{subsec:patient_specific_model}

Tetrahedral meshes were generated for each geometry 0\% RR using the software ANSYS\textregistered  $\;$Meshing\si{^{TM}} (ANSYS, Inc.). A mesh convergence study was performed to verify that the mesh resolution did not significantly affect hemodynamics and stress results. The differences between the results of the selected mesh and a more refined one were less than 1\%. The number of cells in the selected meshes ranged from 1.5M to 2M, depending on each case. 

The simulations were performed using ANSYS\textregistered  $\;$Fluent software (ANSYS, Inc.). The advection and transient schemes were second-order accurate and a $10^{-4}$ convergence criterion was established for the residuals. A coupled scheme was selected as the pressure-velocity coupling algorithm. 

The wall motion generated by the kinematic model described in \S\ref{subsec:wall_motion} was prescribed by applying the dynamic mesh capabilities of ANSYS\textregistered \hspace{0mm}. The node diffusion algorithm was selected as part of the smoothing method, using cell volume as a diffusion function. As both patients suffered from AF, atrial motion was severely reduced and remeshing was not necessary.

As stated in \S\ref{subsec:medical_data}, Doppler velocity measurements in the PV allowed imposing patient-specific boundary conditions at the inlets, while the measurements in the MV gave the possibility of validating the simulation results. An inlet flow and a constant outlet pressure were imposed as boundary conditions. The inlet flow was adjusted to generate the different flow split cases: 59-41\% and 37-63\% LPV-RPV.

Blood was considered a non-Newtonian fluid following Carreau's model \cite{bird87}. The fluid density was assumed to be $\rho = \SI{1050}{kg/m^3}$ and the dynamic viscosity $\mu$ was obtained from the following expression:

\begin{equation} \label{eq:carreau}
\rebuttal{\mu =\mu_{\infty}+\left(\mu_0-\mu_{\infty}\right)\left[1+\left(\lambda\dot\gamma\right)^2\right]^\frac{n-1}{2}},
\end{equation}

where $\dot\gamma$ is the shear rate, $\mu_{\infty}=\SI{0.00345}{\pascal\cdot\second}$ and $\mu_0=\SI{0.056}{\pascal\cdot\second}$ are the infinite and zero shear rate viscosities, $\lambda=\SI{3.31}{\second}$ is the characteristic time, and $n=0.3568$ is the power law index. \rebuttal{These values were taken from previous literature} \cite{CK91}. 

Pressures and velocities were obtained by solving the continuity and Navier-Stokes equations (Eqs. \ref{eq:continuity} and \ref{eq:momentum}, respectively):

\begin{equation} \label{eq:continuity}
\nabla \cdot \textbf{v} = 0\rebuttal{,}
\end{equation}

\begin{equation} \label{eq:momentum}
\frac{\partial{\textbf{v}}}{\partial{t}} + \textbf{v} \cdot \nabla \textbf{v} = - \frac{ \nabla p}{\rho} + \frac{\mu}{\rho} {\nabla}^2 \textbf{v}\rebuttal{,}
\end{equation}\\

where $\textbf{v}$ is the velocity vector and $p$ is the pressure.

A time convergence analysis was performed: the velocity was monitored at several points distributed in the fluid domain, showing that all simulations reached a quasi-periodic regime after four cycles. Therefore, the following analyses shown in \S\ref{sec:results} are based on the results after the fourth cycle. The selected time step was $\Delta t=\SI{0.001}{\second}$, as no variation was detected in the simulation results when it was halved.

\subsection{Hemodynamic indices}
\label{subsec:indices}

\rebuttal{We calculated both age-related and wall shear-related indices. Our approach to calculate the residence time (RT) is based on the Eulerian RT, where an additional advection-diffusion-reaction equation is solved with the flow field. In this case, we extend the RT analysis using the moment generation equation of Sierra-Pallares et al. \cite{Sierra2017} to compute the mean age of the blood and related moments of age, which supply additional information. The moments of age allowed us to delineate stagnant blood regions similarly to our previous work \cite{duenas2021comprehensive, duenas2022morphing, duenas2024reduced, balzotti2023reduced}.} The $m_k$ age moment can be obtained as:

\begin{equation}\label{eq:moment_def}
  m_k = \int_{0}^{\infty}{t^k f(t)\, dt},
\end{equation}

where $f(t)$ is the age distribution of the blood and $t$ is the simulation time.
Fluent User-Defined Functions were used to integrate the moment equations of the age distribution into the solver.

The first moment of the age distribution $m_1$ is the residence time (RT) \rebuttal{of the blood}. The normalized first moment $M_1$ can be obtained as:

  \begin{equation}
    M_1 = \frac{m_1}{\sigma},
  \end{equation}

where $\sigma$ is the standard deviation of $m_1$ and can be obtained as:

\begin{equation}
    \sigma = \sqrt{m_2 - m_1^2}.
\end{equation}

$M_1$ follows a bimodal distribution within the LAA \cite{duenas2021comprehensive}, which allows automatic delimitation of the stagnant volume by selecting a patient-specific threshold. The selected threshold was the minimum probability valley between the two modes.

Some wall shear-based indices were also used for this study, such as the time-averaged wall shear stress (TAWSS) and the oscillatory shear index (OSI). Their formal definition is as follows:

\begin{equation}
\text{TAWSS} = \frac{1}{T} \int_0^T \lvert \text{WSS} \rvert dt\rebuttal{,}
\end{equation}

\begin{equation}
\text{OSI} = \frac{1}{2} \left(1 - \frac{\lvert \int_0^T  \text{WSS } dt \rvert }{\int_0^T \lvert \text{WSS} \rvert dt} \right).
\end{equation}

TAWSS is commonly used as a thrombosis indicator, since low values are associated with endothelial cell damage \cite{chiu2011effects}. OSI is a non-dimensional index that captures oscillations in the wall-shear direction, ranging between 0 and 0.5.

\subsection{Universal left atrial appendage coordinates}

The universal left atrial appendage coordinate (ULAAC) system \cite{duenas2024reduced, rodriguez2024influence} was used to facilitate the visualization and comparison of data from different models. The method was inspired by the universal atrial coordinate (UAC) system of Roney et al. \cite{roney2019universal}, which in turn evolved from the universal ventricular coordinates (UVC) by Bayer et al. \cite{bayer2018universal}. To this end, the LAA surface of both patients was mapped to the unit square by solving two Laplace--Beltrami equations and using some anatomical landmarks as boundary conditions. The Laplace--Beltrami equations were solved using Python LaPy \cite{reuter2006laplace, wachinger2015brainprint}.

\section{Results}\label{sec:results}
Thirty-six atrial simulations were performed since two different flow split ratios (59-41 and 37-63 \% LVP-RPV) were considered for each of the 18 generated cases. All simulations were performed considering a prescribed wall motion and a non-Newtonian model. The cases are distributed as follows:

\begin{itemize}
    \item 10 of these cases correspond to transition AB (Figure \ref{fig:geoms_atb}), summarized in Table \ref{tab:angle-data1}.
    \item 8 of these cases correspond to transition BA (Figure \ref{fig:geoms_bta}), summarized in Table \ref{tab:angle-data2}. 
\end{itemize}

\subsection{Representative atrial flow patterns}
\label{subsec:streamline}

Our main objective was to deepen the understanding of the relationship between PV orientations and LA flow patterns. This phenomenon is important because it affects LAA washing and the risk of medium- and long-term thrombosis. To this end, and as mentioned in \S\ref{sec:methods}, we have modified the PV angles to \rebuttal{obtain gradual transitions between the two opposite types of atrial flow patterns A and B} \cite{rodriguez2024influence}. In an effort to isolate the influence of the PV orientations, we have kept the rest of the parameters constant for each anatomy: cardiac output, atrial motion, LAA volume, morphology, etc.

\begin{figure*}[t!]
    \centering
   \begin{subfigure}[t]{0.63\textwidth}
        \centering
        \includegraphics[width=\linewidth]{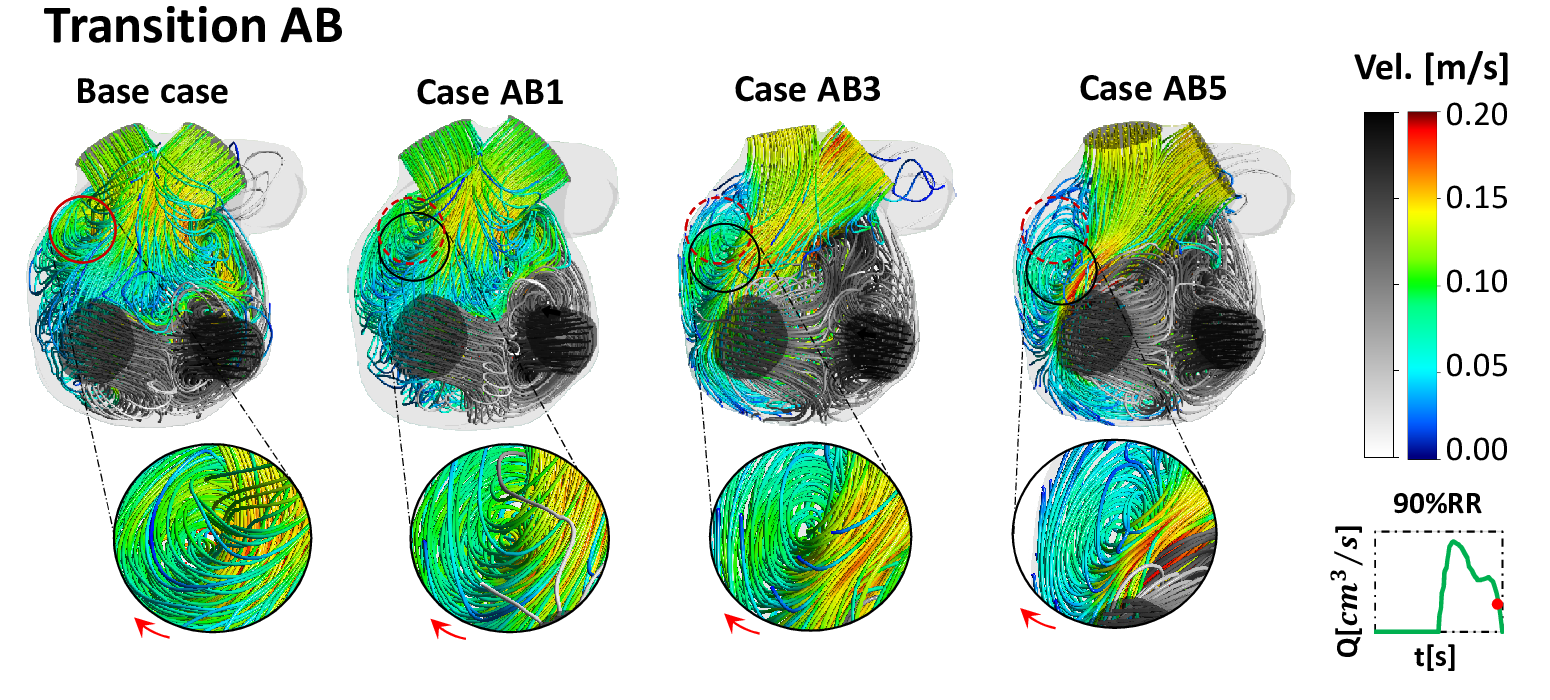}
        \caption{}
          \label{subfig:streamlines_atb1}
    \end{subfigure}
    \begin{subfigure}[t]{0.63\textwidth}
        \centering
        \includegraphics[width=\linewidth]{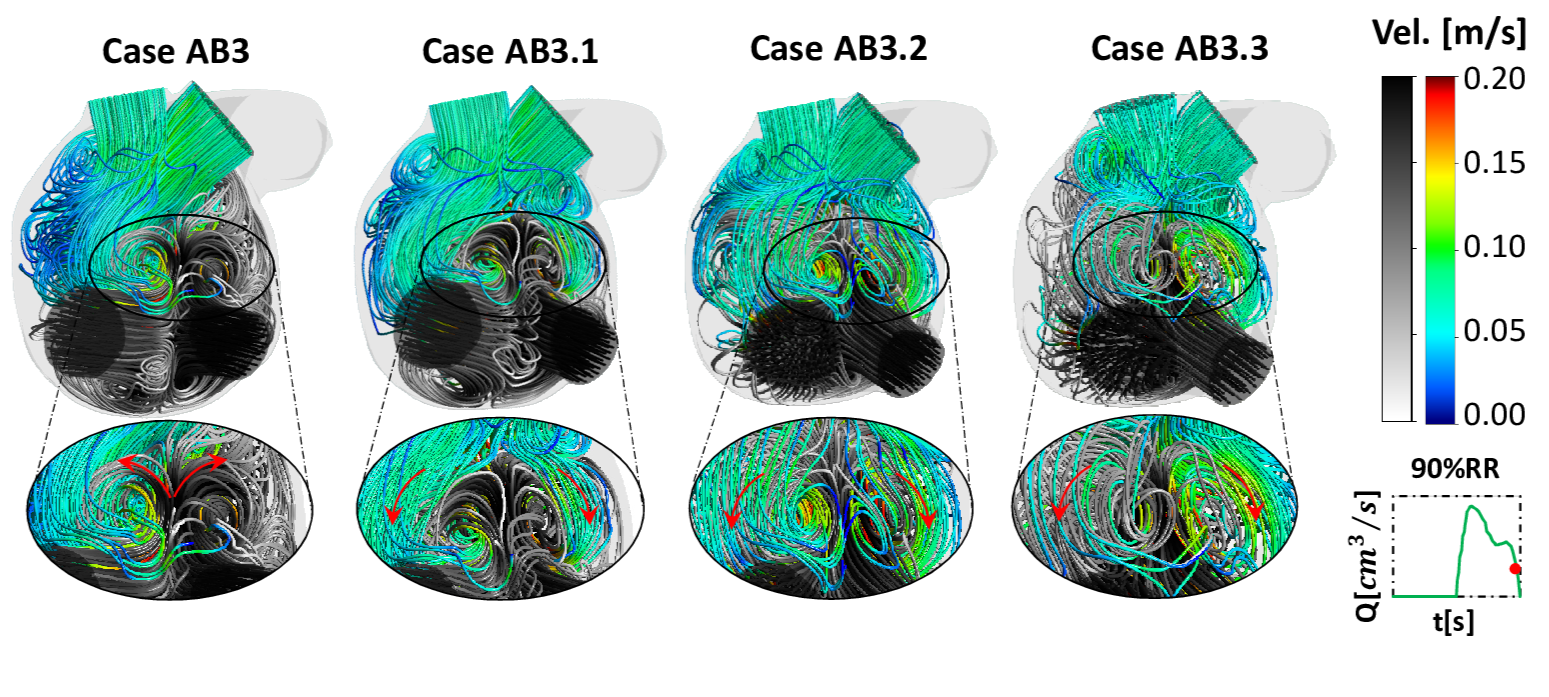}
        \caption{}
          \label{subfig:streamlines_atb2}
    \end{subfigure}   
             \begin{subfigure}[t]{0.63\textwidth}
        \centering
        \includegraphics[width=\linewidth]{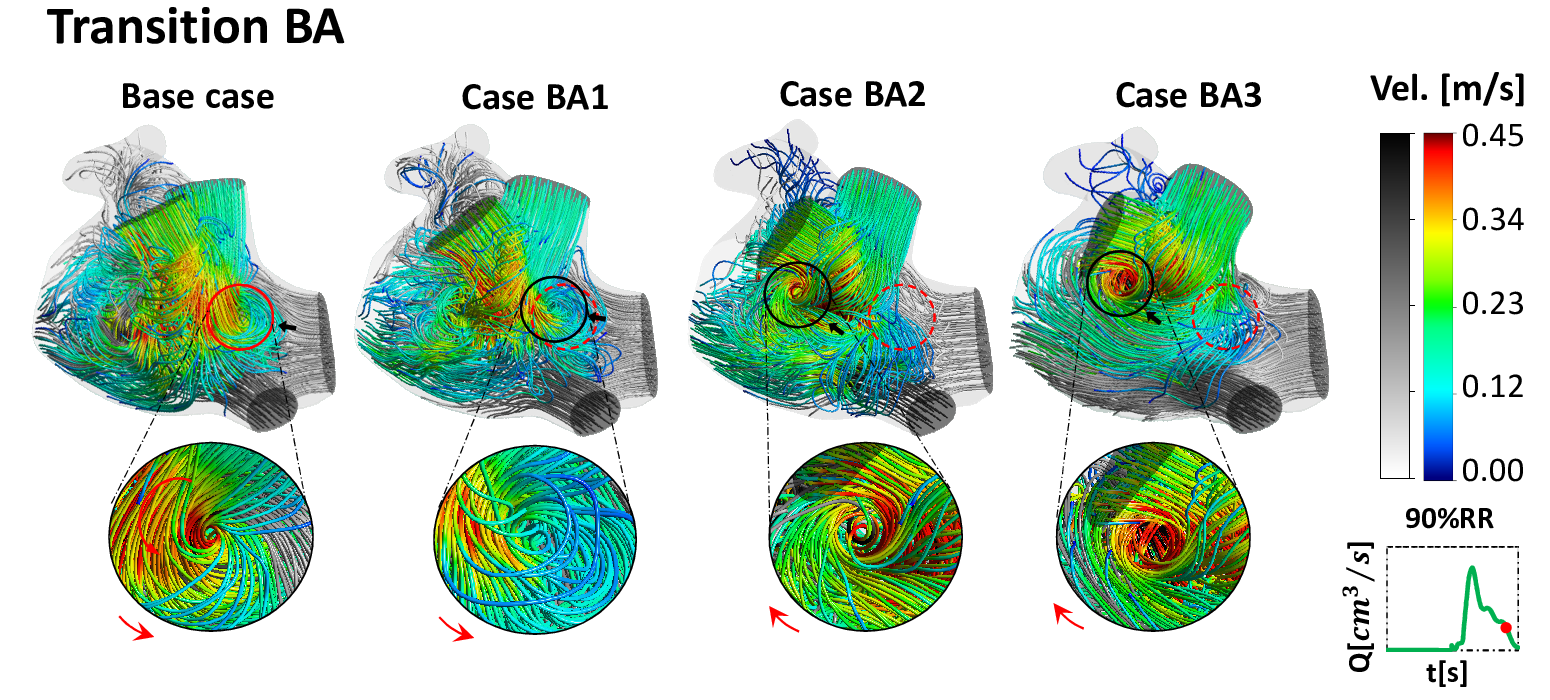}
        \caption{}
          \label{subfig:streamlines_bta1}
    \end{subfigure}
    \begin{subfigure}[t]{0.63\textwidth}
        \centering
        \includegraphics[width=\linewidth]{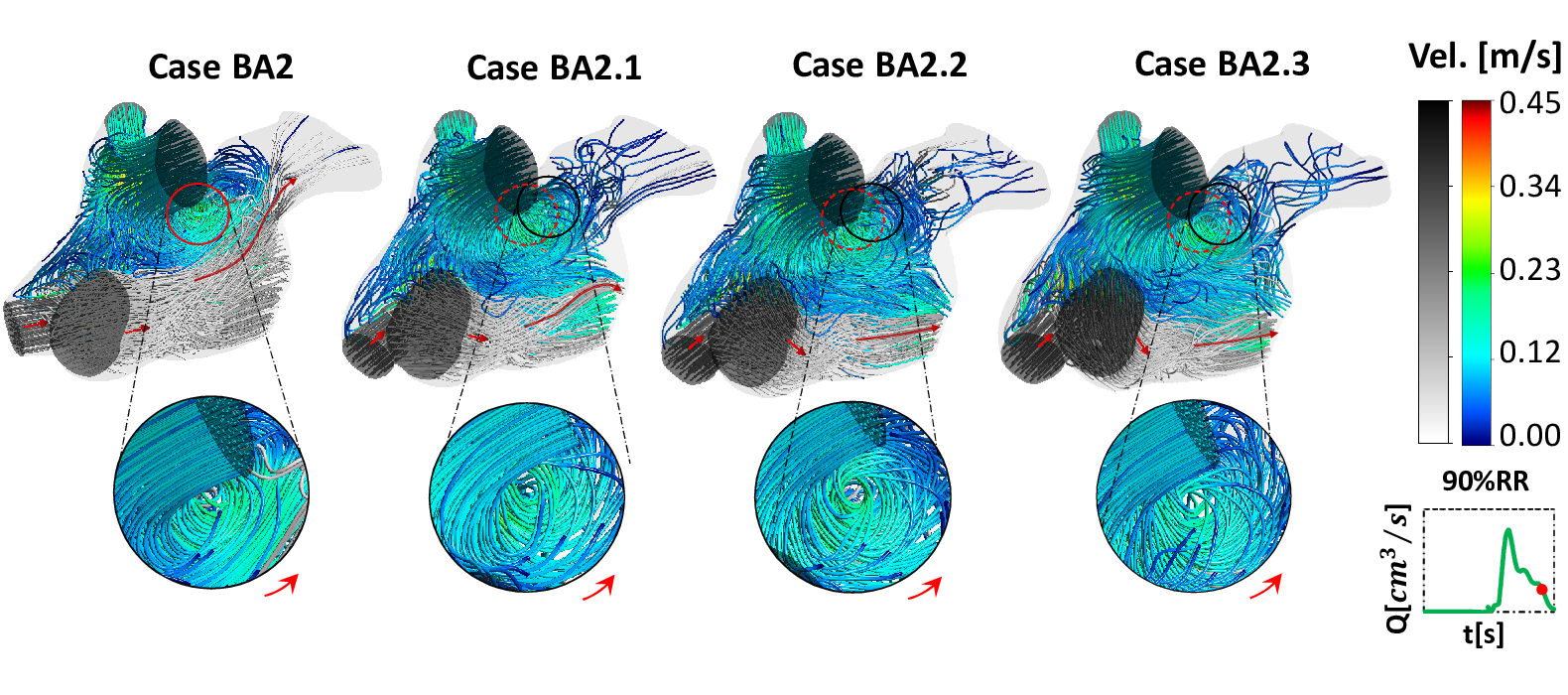}
        \caption{}
          \label{subfig:streamlines_bta2}
    \end{subfigure}    

\caption{\rebuttal{Blood flow streamlines at 90\%RR for transitions AB (a,b) and BA (c,d). The color (gray) scale represents the velocity magnitude of the flow entering LA through the LPV (RPV). Red circles represent the position of the center of the circulatory flow in the base case. In contrast, the black ones represent the position when the PV orientation is modified. Subfigures (a,c) and (b,d) correspond to a 59-41\%  and 37-63\% LPV-RPV flow split ratio, respectively. LA has been rotated to facilitate flow visualization, and zooms were added. Arrows indicate the direction of rotation. Two animations of this figure are included (Supplementary Data A.2.1 and A.2.2).}}
\label{fig:velocity_streamlines}
\end{figure*}

The first step was to identify the atrial flow patterns for both transitions, which was achieved by analyzing the blood streamlines. Figure \ref{fig:velocity_streamlines} represents the blood streamlines when the flow structures are fully developed in 90\%RR. 

Regarding Transition AB, as stated previously, the LPV angles were modified so that the LPV were progressively closed. During the modification of the LPV, the center of the main circulatory flow moved from the LPV towards the LA roof. This can be seen in the streamlines for the cases AB1-AB5 in Figure \ref{subfig:streamlines_atb1} \rebuttal{and Table \ref{tab_distance}}, where the flow-split ratio 59-41\% was displayed to improve visualization of the main circulatory flow. These cases present a negligible interaction between LPV and RPV flows (colored LPV streamlines did not interact with gray RPV streamlines, Figure \ref{subfig:streamlines_atb1}), making it necessary to modify the RPV orientations to complete the AB transformation.

Case AB3 was selected as the starting point for the RPV transformation, as it presented more streamlines entering the LAA and, thus, better washing. As described above, the RPV transformation implies providing the RPV a caudal orientation in the axial plane while closing them in the coronal plane. 
An increasing interaction can be seen in the streamlines in Figure \ref{subfig:streamlines_atb2} between the RPV and LPV flows for cases AB3.1-3.3. Simulations with a flow-split ratio of 37-63\% are shown for these cases, since this LPV-RPV flow interaction does not occur for 59-41\%. The transformation from a type A to a type B atrial flow pattern is now complete.

As stated above, the LPV were opened in the coronal plane in the transition BA. Figure \ref{subfig:streamlines_bta1} \rebuttal{and Table \ref{tab_distance}} shows how, for \rebuttal{BA1-3 cases}, the main circulatory flow changes its rotation direction and progressively moves toward the exit of the LPV. Similarly to Transition AB, 59-41\% cases are shown to improve visualization of the main circulatory flow. The streamlines show how the washing of the LAA gradually becomes more governed by the LPV flow. However, pointing the RPV flow directly towards the MV is still necessary to complete the transformation to type A.

Case BA2 was selected as the starting point for the RPV transformation, as it presented a well-defined circulatory flow at the exit of the LPV: the center of the circulatory flow is clearly visible, and the RPV flow just surrounds it. After surrounding the center of circulatory flow, the RPV streamlines enter \rebuttal{the} LAA (see case BA2 in Figure \ref{subfig:streamlines_bta2} where only gray streamlines can be observed in LAA). For BA2.1-2.3 cases, the superior RPV were gradually oriented cranially, while the inferior RPV were oriented caudally. Figure \ref{subfig:streamlines_bta2} also shows how RPV flow drives the main circulatory vortex near the ostium, improving LAA washing. Case BA2.3 shows how LPV flow is responsible for LAA washout in 37-63\% (only color streamlines can be observed in LAA). This reproduces the typical behavior of a type A atrial flow pattern, so the BA transformation is complete. 

\begin{table}[htbp]
\centering
\caption{\rebuttal{Distance between the center of the ostium and the center of the main circulatory flow for the transitions AB and BA  corresponding to a 59-41\% LPV-RPV flow split ratio.}}
\color{black}
\begin{tabular}{lccl}
  & \textbf{Case} & \textbf{Distance [m]} &   \\ 
\cline{1-3} \multirow{4}{*}{\textbf{Transition AB}} & Base   & 0.036    &   \\ \cline{2-3}   & 1    & 0.039    &   \\ \cline{2-3}   & 3             & 0.040   &   \\ \cline{2-3}   & 5    & 0.045    &   \\ 
\cline{1-3} \multirow{4}{*}{\textbf{Transition BA}} & Base   & 0.057        &   \\ \cline{2-3}  & 1  & 0.048   &   \\ \cline{2-3}  & 2             & 0.043    &   \\ \cline{2-3} & 3    & 0.037  &  \\
\cline{2-3}
\end{tabular}
\label{tab_distance}
\end{table}

\subsection{Analysis of the vortex structures}
\label{subsec:vor_struc}

\begin{figure*}[t!]
    \centering
   \begin{subfigure}[t]{0.48\textwidth}
        \centering
        \includegraphics[width=\linewidth]{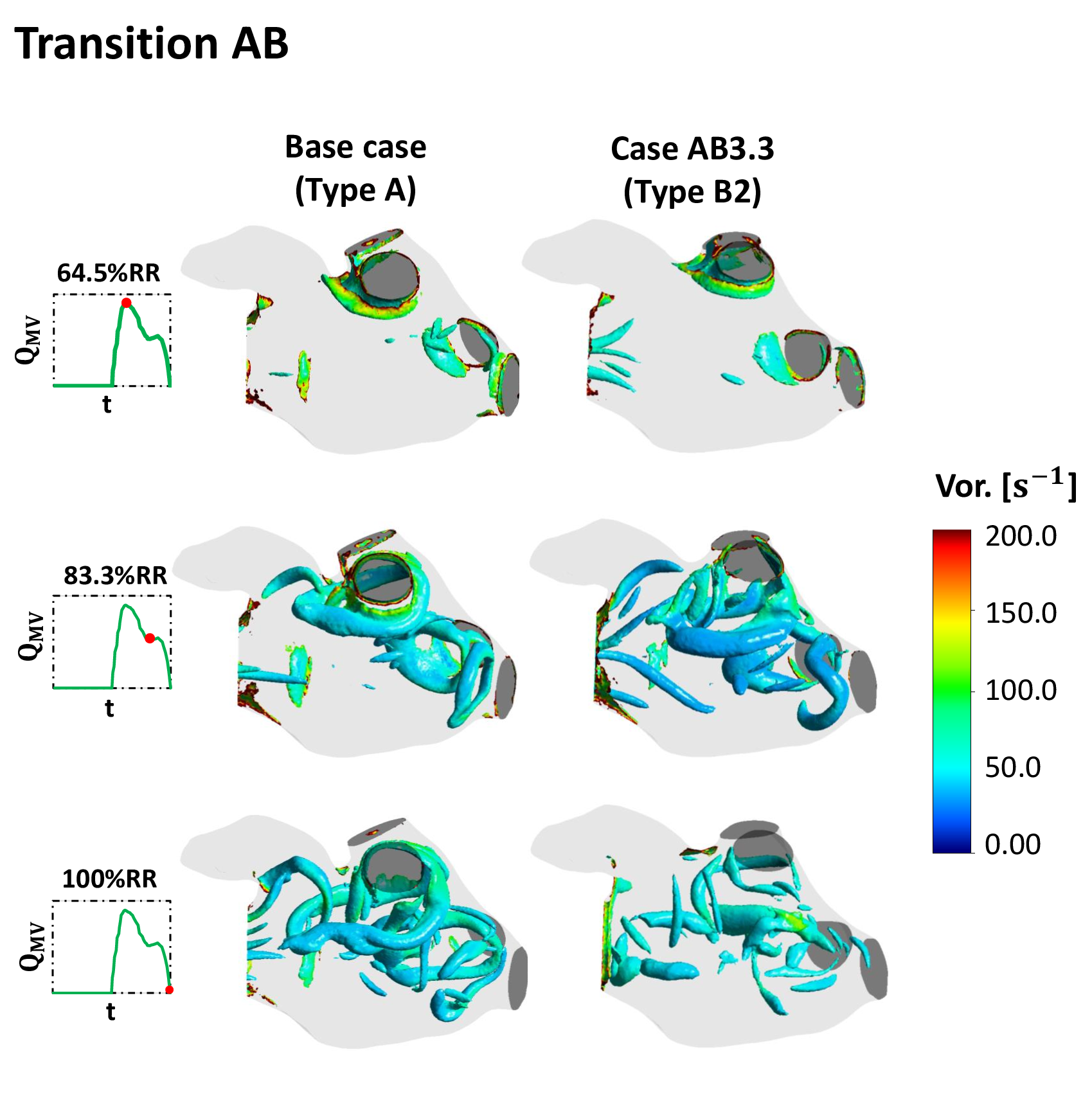}
        \caption{}
          \label{subfig:vor_vortex_atb}
    \end{subfigure}
    \begin{subfigure}[t]{0.48\textwidth}
        \centering
        \includegraphics[width=\linewidth]{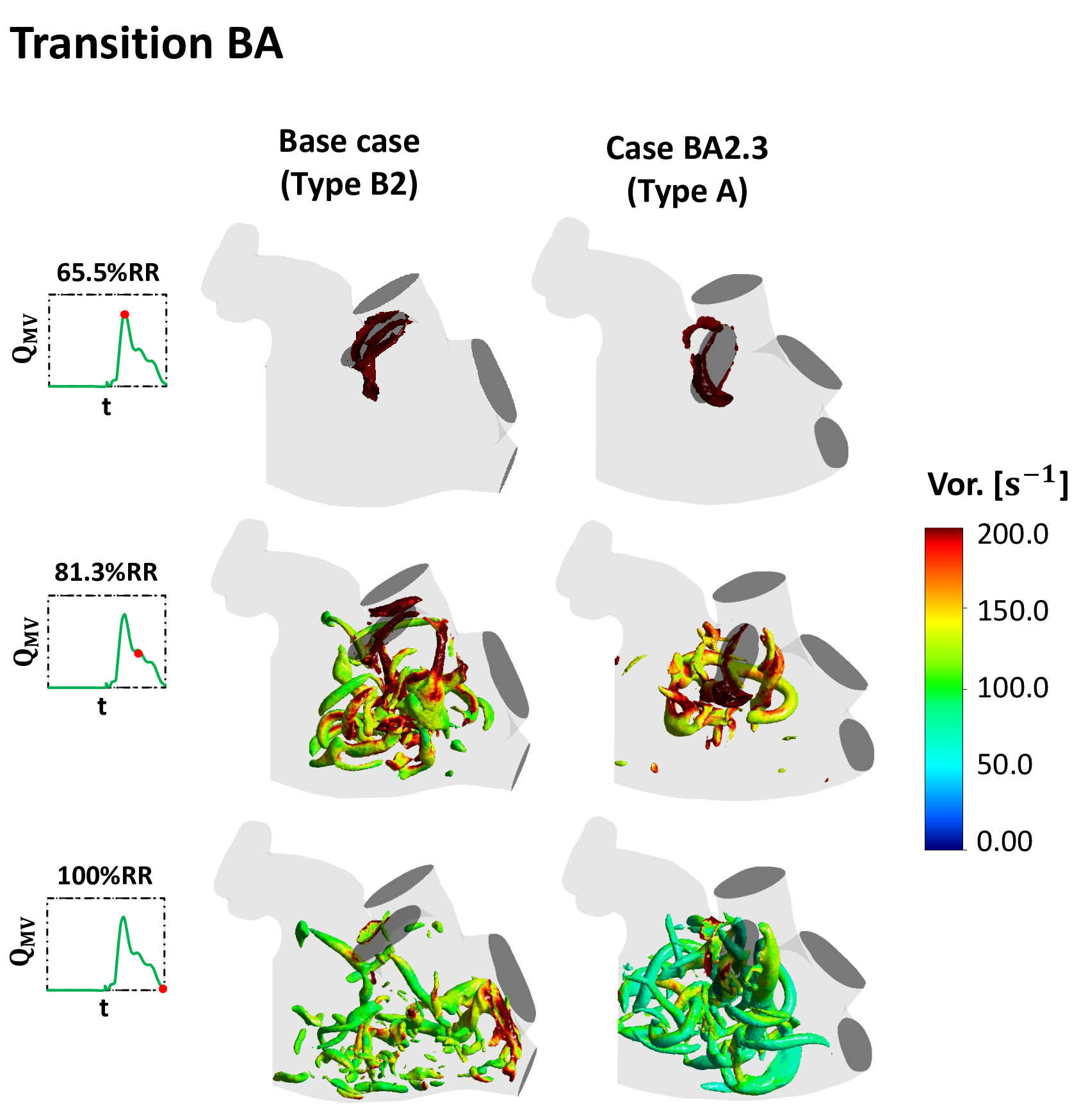}
        \caption{}
          \label{subfig:vor_vortex_bta}
    \end{subfigure}    
    \begin{subfigure}[t]{0.48\textwidth}
        \centering
        \includegraphics[width=\linewidth]{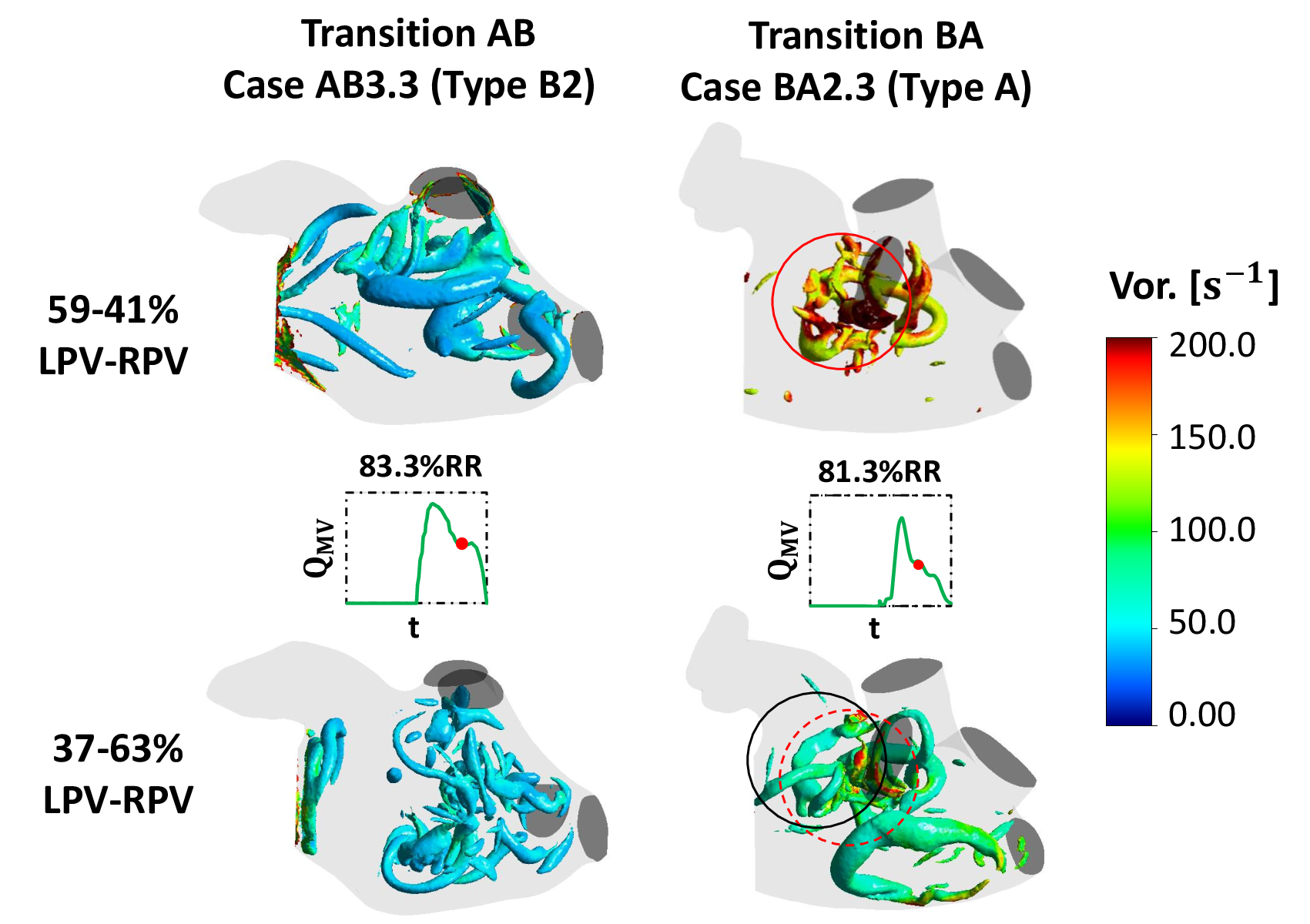}
        \caption{}
          \label{subfig:vor_vortex_comp}
    \end{subfigure}    
\caption{\rebuttal{Vortex flow structures during systole for transitions AB (a) and BA (b), respectively. (c) Evolution of the main vortex structure when the flow split ratio changes from 59-41\%LPV-RPV to 37-63\%LPV-RPV. Flow structures were generated according to the $q_2$ criterion ($q_2 = 0.2$) and colored according to their vorticity value. The red circles indicate the initial position of the vortex, while the black circle marks the final position. Two animations of this figure are included (Supplementary Data A.3.1 and A.3.2).}}
\label{fig:vor_vortex_evo}
\end{figure*}

The results of the previous section can be confirmed by analyzing the atrial vortex structures. Figure \ref{fig:vor_vortex_evo} shows the vortex structures of the AB and BA transitions at different \rebuttal{times} of the cardiac cycle. The chosen instants correspond to diastole, as during systole, when the MV is closed, it is impossible to detect a main vortex structure. The first instant corresponds to the maximum blood flow through the PVs (64.5\%RR). The second corresponds to the instant when the maximum vortex structure development is achieved (81.3\%RR). Finally, the third corresponds to the instant of the MV closure (100\%RR).

Figure \ref{subfig:vor_vortex_atb} shows the comparison of the vortex structures between the initial (Base case - Type A flow) and the final morphology (Case AB3.3 - Type B flow) for the AB transition. The vortex structures are not visible for both cases in 64.5\% RR, as the flow patterns are not fully developed. The maximum vorticity values are reached near the PV. The main vortex structures are fully developed at 83.3\%RR, \rebuttal{varying their location from the LPV (Base case - Type A flow) to the LA roof (Case AB3.3 - Type B flow). The vorticity value is also reduced from 64.1 s$^{-1}$ to 32.3 s$^{-1}$. Regarding MV closure, the vortex structures of the Base case are mainly conserved, while they are substantially modified for Case AB3.3, where several vortex structures with lower vorticities are dissipated.}

Regarding the BA transition, Figure \ref{subfig:vor_vortex_bta} also compares the vortex structures between the initial \rebuttal{(Base case, Type B flow)} and the final morphology (Case BA2.3 - Type A flow). Following the same criterion as in the AB transition, three instants were selected when MV was opened. The high inflow at 65.5\%RR marks the beginning of the formation of vortex structures. As in transition AB, \rebuttal{the} vorticity values peak at this instant. The vorticity values are significantly higher for transition BA ($\sim$200 s$^{-1}$). The complete development of the main vortex structures occurs at 81.3\%RR. At this stage, the vortex structures are concentrated near the LPV (case BA2.3), following the behavior of a type A atrium (Base case in Figure \ref{subfig:vor_vortex_atb}). In contrast, the base case shows a hectic flow structure close to the LA roof, following the behavior of a type B atrium. This structure is further fractured when the MV is closed, similar to Case AB3.3 (Figure \ref{subfig:vor_vortex_atb}). On the other hand, Case BA2.3 concentrates its vortex structures around LPV, its vorticity value being similar to the base case in Figure \ref{subfig:vor_vortex_atb}.

Finally, Figure \ref{subfig:vor_vortex_comp} shows the influence of the flow split ratio when the main vortex structures are fully developed (83.3\% and 81.3\%RR for transitions AB and BA, respectively). For transition AB, the vortex structures are disordered by the collision of the RPV flow with the main vortex of the LPV when the flow split ratio is changed to 37-63\%LPV-RPV. Regarding the BA transition, the main vortex is pushed towards the LAA ostium by the RPV flow when the flow split ratio is changed to 37-63\%LPV-RPV.

\subsection{Left atrial appendage residence time}
\label{subsec:LAA_RT}
\begin{figure*}[t]
 \centering
   \begin{subfigure}[t]{0.95\textwidth}
        \centering
        \includegraphics[width=\linewidth]{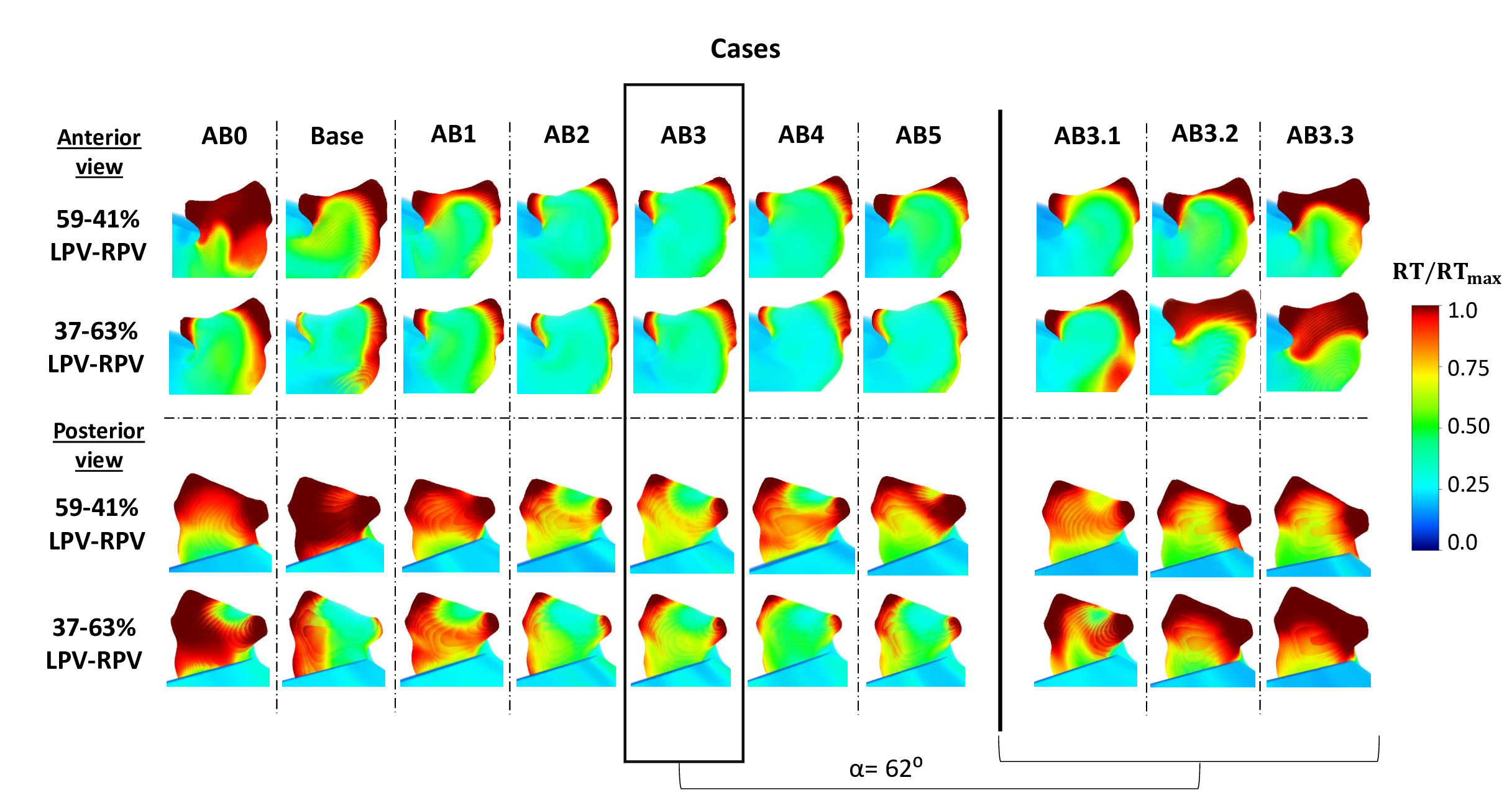}
        \caption{}
          \label{subfig:rt_atb}
    \end{subfigure}
    \begin{subfigure}[t]{0.95\textwidth}
        \centering
        \includegraphics[width=\linewidth]{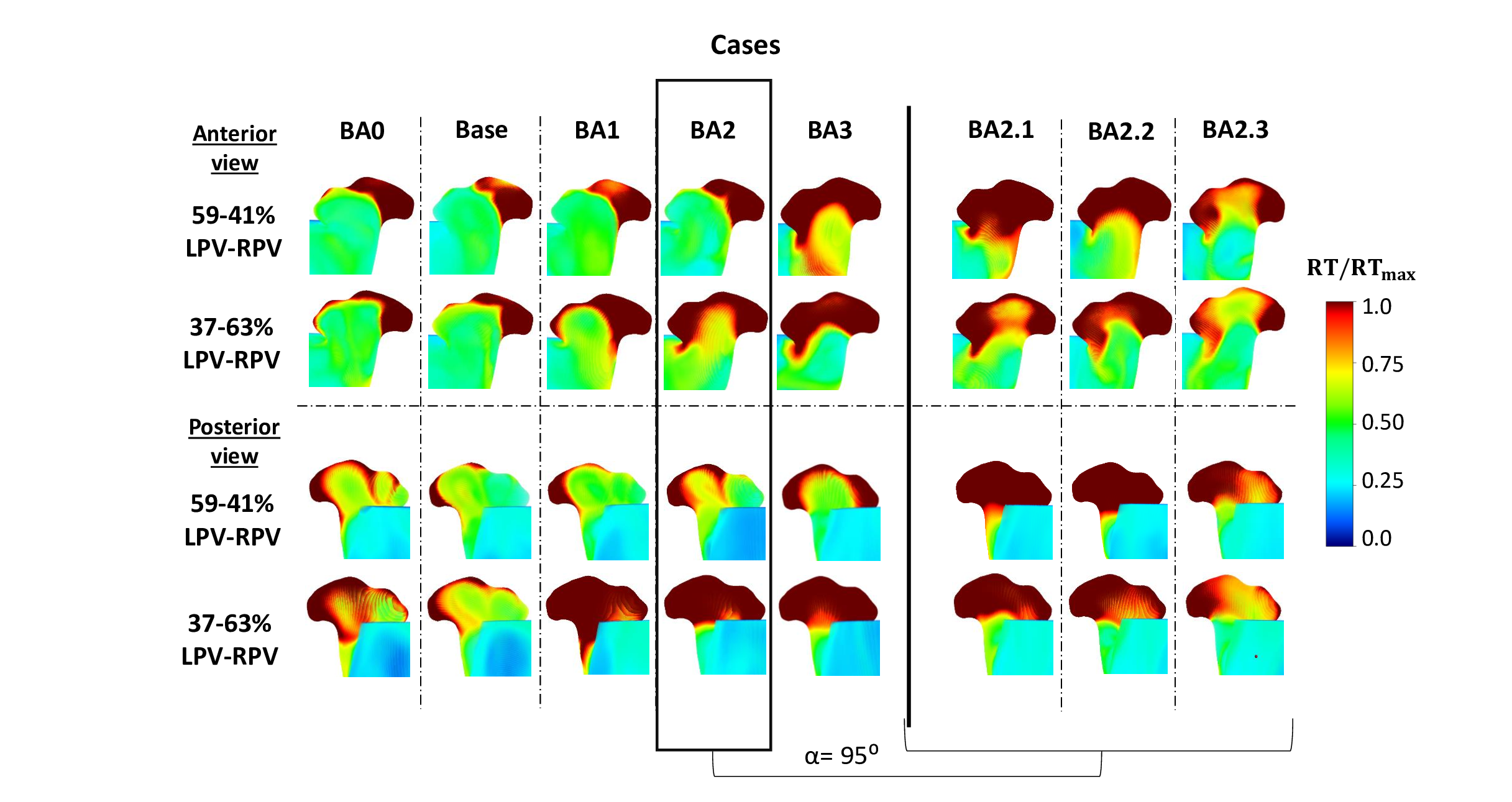}
        \caption{}
          \label{subfig:rt_bta}
    \end{subfigure}    
\caption{Stagnant volume for the different flow split ratios (rows), and PV configurations (columns) for (a) transition AB, and (b) transition BA. The cases are grouped according to the PVs that were modified (LPVs on the left and RPVs on the right).}
\label{fig:LAA_RT_by_transformation}
\end{figure*}

Streamlines and vortex structures have shown that LA flow patterns can be substantially modified when changing the PV orientation. As LA flow patterns govern LAA washing, these changes in the PV orientation should lead to alterations in the RT within the LAA. Figure \ref{fig:LAA_RT_by_transformation} shows $\text{RT}/\text{RT}_{\text{max}}$ contour evolution for all PV orientations (columns) for the flow splits under consideration. For each patient, RT is normalized with its maximum value, RT$_{\mbox{max}}$.
Figure \ref{subfig:rt_atb} depicts $\text{RT}/\text{RT}_{\text{max}}$ evolution for the transition AB. When the LPV orientation is modified (AB0-AB5) while keeping the flow split at 59-41\%LPV-RPV (see \S\ref{subsec:morphological_descriptors}), $\text{RT}/\text{RT}_{\text{max}}$ gradually decreases until it reaches a minimum in Case AB3, growing afterward. $\text{RT}/\text{RT}_{\text{max}}$ increases in cases AB4 and AB5 because $\Phi_1$ exceeds 180$^{\circ}$ and the original flow direction is reversed. A similar trend is observed when the flow split ratio is changed to 37-63\%, but $\text{RT}/\text{RT}_{\text{max}}$ decreases with respect to 59-41\%. Therefore, as described in \S\ref{subsec:streamline}, the modification of the RPV orientation is necessary to complete the transition AB (see the cases AB3.1, AB3.2 and AB3.3 in Figure \ref{subfig:rt_atb}).

Figure \ref{subfig:rt_bta} shows the evolution of the contour $\text{RT}/\text{RT}_{\text{max}}$ for the BA transition. In this transition, the values of $\text{RT}/\text{RT}_{\text{max}}$ increase when the LPV orientations are modified (BA0-BA3). This increase is more noticeable for the 37-63\% flow split. As explained above, a change in the RPV orientation is required to reverse this increase and complete the transition BA (see cases BA2.1, BA2.2, and BA3.3 in Figure \ref{subfig:rt_bta}).

\subsection{Stagnant blood volume evolution}
\label{stag_vol_evolution}
\begin{figure*}[t!]
    \centering
    \begin{subfigure}[t]{0.48\textwidth}
        \centering
        \includegraphics[width=\linewidth]{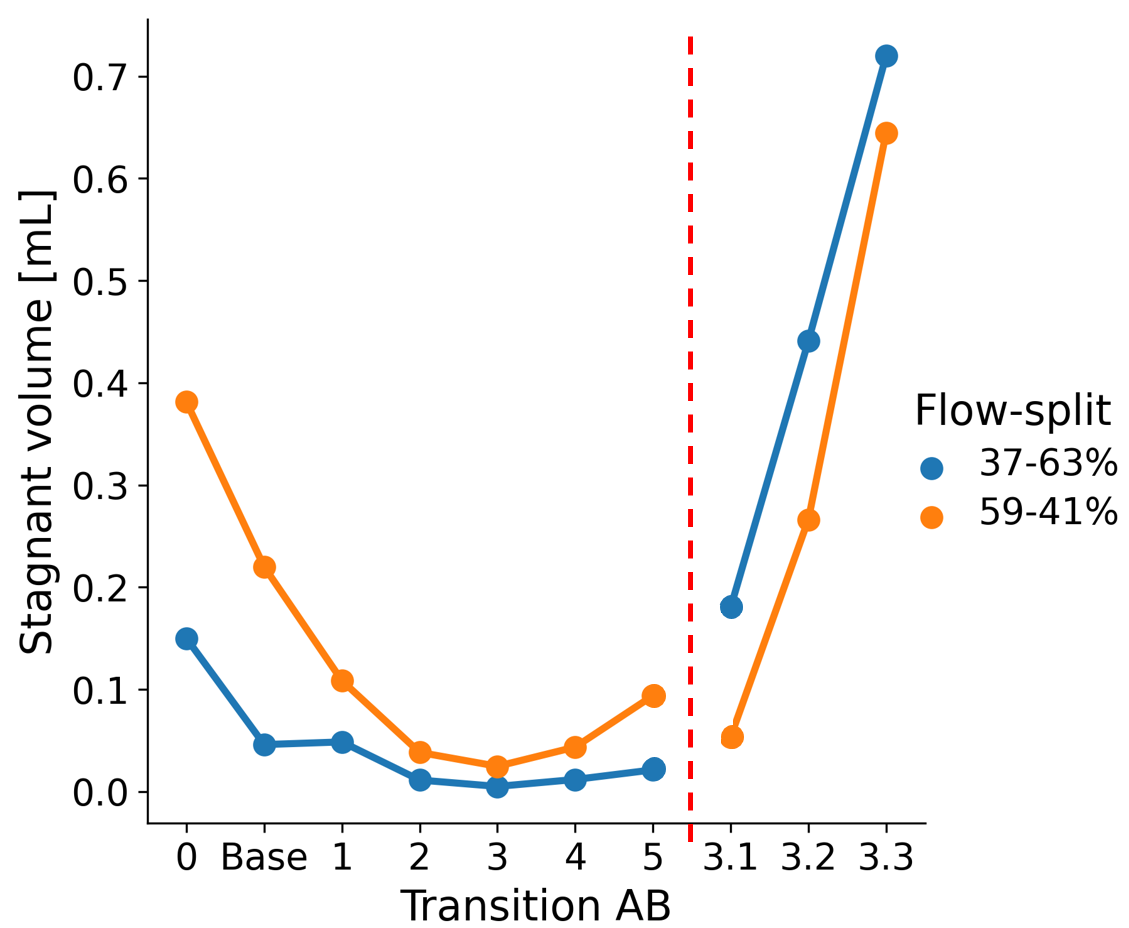}
        \caption{}
        \label{subfig:stagnant_volume_atb}
    \end{subfigure}
    \begin{subfigure}[t]{0.48\textwidth}
        \centering
        \includegraphics[width=\linewidth]{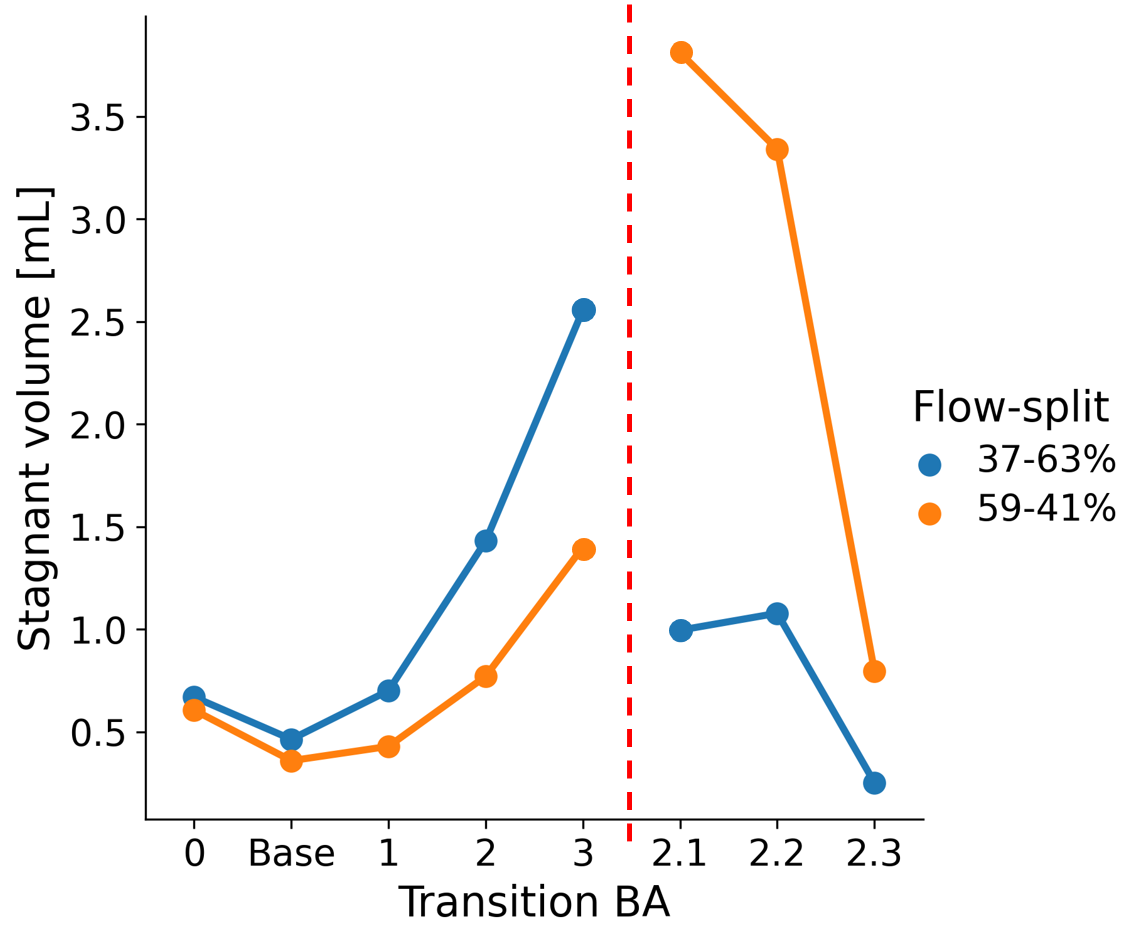}
        \caption{}
          \label{subfig:stagnant_volume_bta}
    \end{subfigure}
   
\caption{Evolution of the stagnant volume for the simulated cases for the transitions AB (a) and BA (b). The dashed vertical line separates the cases in which the LPV or the RPV are modified, respectively. Orange and blue lines represent the different flow split ratios under consideration.}
\label{fig:stagnant_volume_evolution}
\end{figure*}

An interesting parameter to delineate the stagnant region and thus determine the risk of stasis in the LAA is the stagnant volume. It was calculated as described in \S \ref{subsec:indices}, and its evolution is depicted in Figure \ref{fig:stagnant_volume_evolution} \rebuttal{and Tables \ref{tab:stagnant_evolution_atb} and \ref{tab:stagnant_evolution_bta}}. Figures \ref{subfig:stagnant_volume_atb} and \ref{subfig:stagnant_volume_bta} graph the stagnant volume trend for the AB and BA transitions, respectively. The stagnant volume follows a trend similar to $\text{RT}/\text{RT}_{\text{max}}$. For the transition AB, it decreases when LPV are closed in the coronal plane (cranial orientation in the axial plane), reaches a minimum in case AB3, and increases slightly for cases AB4-AB5 \rebuttal{(Table \ref{tab:stagnant_evolution_atb})}. Then, as mentioned, the RPV are modified (caudal orientation in the axial plane) in the AB3.1-AB3.3 cases to achieve type B flow. This is reflected in Figure \ref{subfig:stagnant_volume_atb} \rebuttal{and Table \ref{tab:stagnant_evolution_atb}}, where the stagnant volume \rebuttal{abruptly increases.}

Regarding the BA transition, the LPV are opened in the coronal plane (the LSPV are caudally oriented and the LIPV are cranially oriented in the axial plane). Figure \ref{subfig:stagnant_volume_bta} \rebuttal{and Table \ref{tab:stagnant_evolution_bta}} depicts the evolution of the stagnant volume: In this transition, the minimal stagnant volume is found for the base case, as any modification in the LPV orientation leads to an increase. The highest value is reached for case BA3. When the RPV are modified (the RSPV cranially oriented and the RIPV caudally oriented), the stagnant volume drastically increases in case BA2.1 for the flow split ratio 59-41\%LPV-RPV, slightly decreasing for 37-63\%. The transition to type A is confirmed when RPVs are further modified in the BA2.2-BA2.3 cases, as the stagnant volume decreases for the 59-41\%LPV-RPV flow split ratio (it increases slightly and then decreases for the 37-63\% LPV-RPV flow split ratio (\rebuttal{Table \ref{tab:stagnant_evolution_bta}}).

\begin{table}
\centering
\caption{\rebuttal{SBV variation for transition AB. The values are depicted in Figure \ref{subfig:stagnant_volume_atb}.}}
\begin{adjustbox}{max width=8.5cm}
\color{black}
\begin{tabular}{cccc}
                                & Case                  & \%LPV-RPV & SBV~[mL]~  \\ 
\hline
\multirow{20}{*}{Transition AB} & \multirow{2}{*}{0}    & 59-41     & 0.38       \\
                                &                       & 37-63     & 0.15       \\ 
\cline{2-4}
                                & \multirow{2}{*}{Base} & 59-41     & 0.22       \\
                                &                       & 37-63     & 0.05       \\ 
\cline{2-4}
                                & \multirow{2}{*}{1}    & 59-41     & 0.11       \\
                                &                       & 37-63     & 0.05       \\ 
\cline{2-4}
                                & \multirow{2}{*}{2}    & 59-41     & 0.04       \\
                                &                       & 37-63     & 0.01       \\ 
\cline{2-4}
                                & \multirow{2}{*}{3}    & 59-41     & 0.02       \\
                                &                       & 37-63     & 0.01       \\ 
\cline{2-4}
                                & \multirow{2}{*}{4}    & 59-41     & 0.04       \\
                                &                       & 37-63     & 0.01       \\ 
\cline{2-4}
                                & \multirow{2}{*}{5}    & 59-41     & 0.09       \\
                                &                       & 37-63     & 0.02       \\ 
\cline{2-4}
                                & \multirow{2}{*}{3.1}  & 59-41     & 0.05       \\
                                &                       & 37-63     & 0.18       \\ 
\cline{2-4}
                                & \multirow{2}{*}{3.2}  & 59-41     & 0.27       \\
                                &                       & 37-63     & 0.44       \\ 
\cline{2-4}
                                & \multirow{2}{*}{3.3}  & 59-41     & 0.64       \\
                                &                       & 37-63     & 0.72       \\
\hline
\end{tabular}
\end{adjustbox}
\label{tab:stagnant_evolution_atb}
\end{table}

\begin{table}[htbp]
\centering
\caption{\rebuttal{SBV variation for transition BA. The values are depicted in Figure \ref{subfig:stagnant_volume_bta}.}}
\begin{adjustbox}{max width=8.5cm}
\color{black}
\begin{tabular}{cccc}
                                & Case                  & \%LPV-RPV & SBV~[mL]~  \\ 
\hline
\multirow{16}{*}{Transition BA} & \multirow{2}{*}{0}    & 59-41     & 0.61       \\
                                &                       & 37-63     & 0.67       \\ 
\cline{2-4}
                                & \multirow{2}{*}{Base} & 59-41     & 0.36       \\
                                &                       & 37-63     & 0.46       \\ 
\cline{2-4}
                                & \multirow{2}{*}{1}    & 59-41     & 0.43       \\
                                &                       & 37-63     & 0.70       \\ 
\cline{2-4}
                                & \multirow{2}{*}{2}    & 59-41     & 0.77       \\
                                &                       & 37-63     & 1.43       \\ 
\cline{2-4}
                                & \multirow{2}{*}{3}    & 59-41     & 1.40       \\
                                &                       & 37-63     & 2.56       \\ 
\cline{2-4}
                                & \multirow{2}{*}{2.1}  & 59-41     & 3.81       \\
                                &                       & 37-63     & 1.00       \\ 
\cline{2-4}
                                & \multirow{2}{*}{2.2}  & 59-41     & 3.34       \\
                                &                       & 37-63     & 1.08       \\ 
\cline{2-4}
                                & \multirow{2}{*}{2.3}  & 59-41     & 0.80       \\
                                &                       & 37-63     & 0.25       \\
\hline
\end{tabular}
\end{adjustbox}
\label{tab:stagnant_evolution_bta}
\end{table}

\subsection{Hemodynamic maps}
\label{subsec:hemodynamic_maps}

\begin{figure*}[t]
 \centering
\centerline{\includegraphics[width=0.99\linewidth]{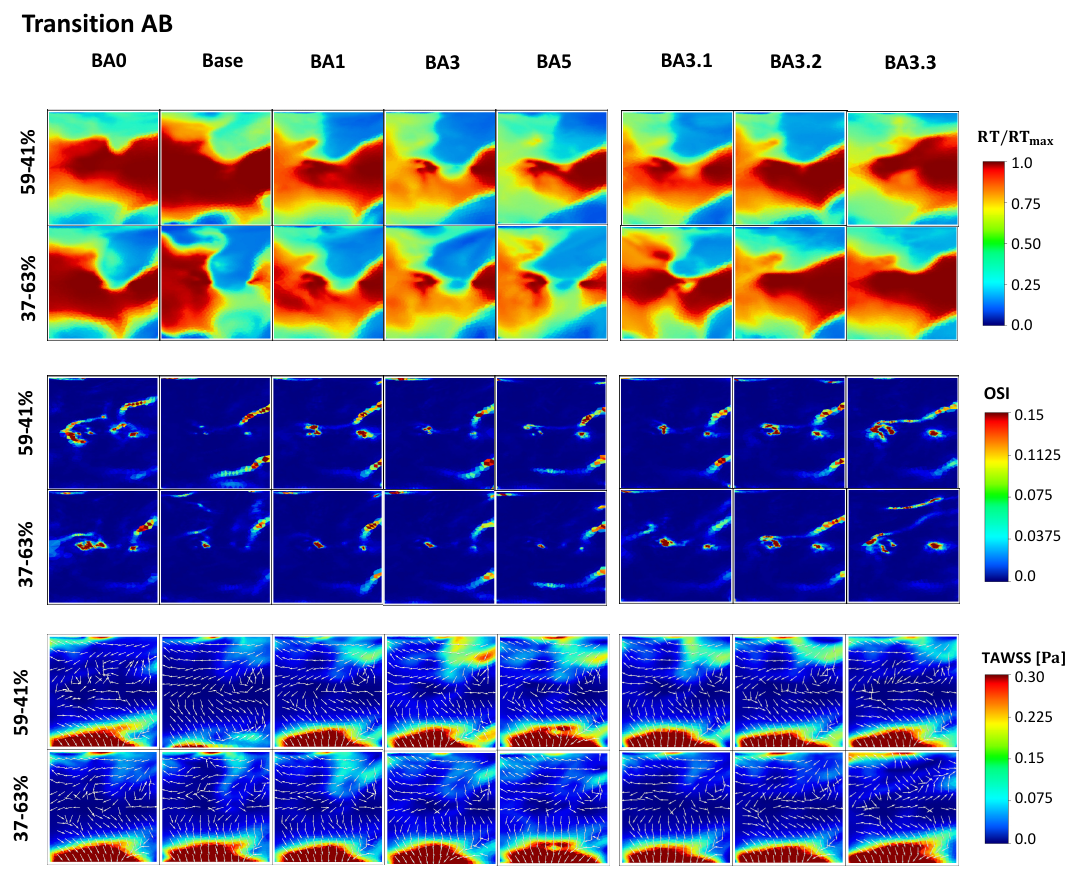}}
\caption{\srebuttal{ULAAC projections of $\text{RT}/\text{RT}_{\text{max}}$, OSI, and TAWSS for the transitions AB}. The columns show the evolution of the contours through the different PV configurations, while the rows show the changes due to the different flow split ratios considered. \srebuttal{The direction of the time-averaged WSS vector is displayed along with the TAWSS contours.}}
\label{fig:ulaac_atb}
\end{figure*}

\begin{figure*}[t]
\centerline{\includegraphics[width=0.99\linewidth]{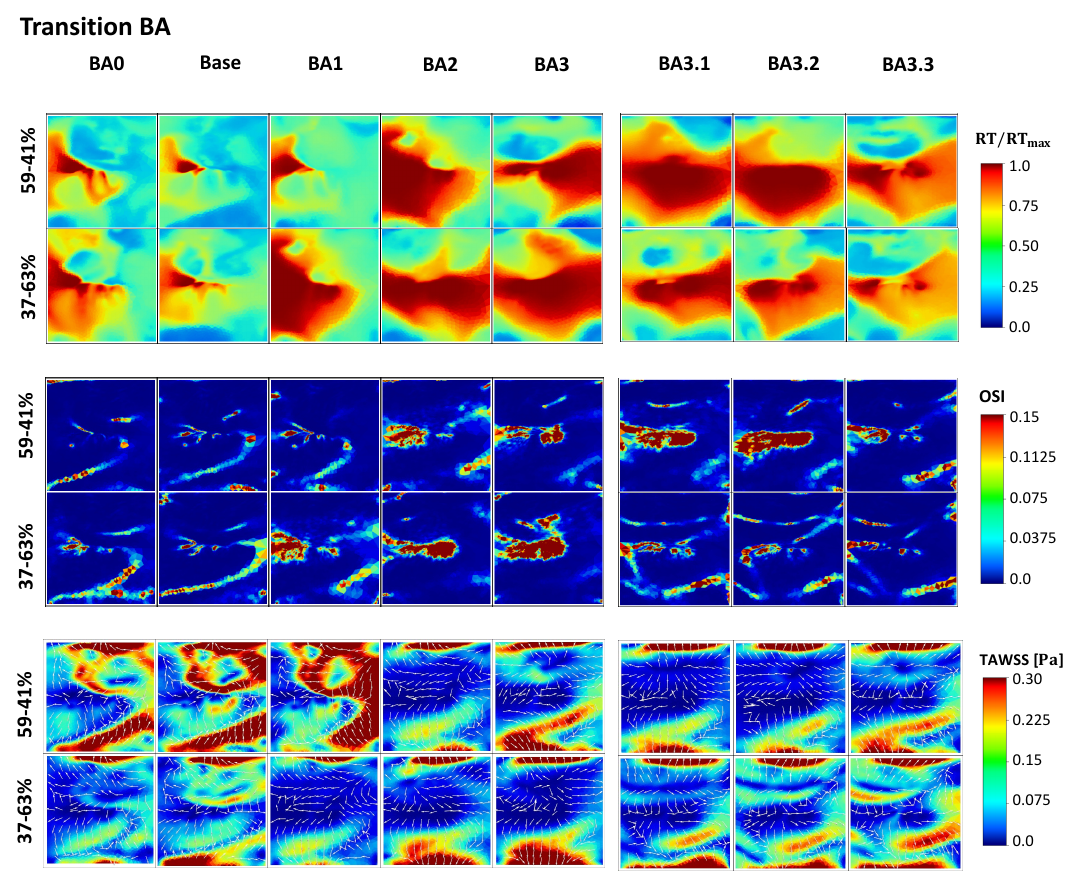}}
\caption{\srebuttal{ULAAC projections of $\text{RT}/\text{RT}_{\text{max}}$, OSI, and TAWSS for the transitions BA}. The columns show the evolution of the contours through the different PV configurations, while the rows show the changes due to the different flow split ratios considered.\srebuttal{The direction of the time-averaged WSS vector is displayed along with the TAWSS contours.}}
\label{fig:ulaac_bta}
\end{figure*}

\srebuttal{Figures \ref{fig:ulaac_atb} and \ref{fig:ulaac_bta}} shows the hemodynamic maps for $\text{RT}/\text{RT}_{\text{max}}$, TAWSS and OSI, where RT was normalized using the duration of the cardiac cycle. The hemodynamic indices were projected into the ULAAC system \cite{duenas2024reduced, rodriguez2024influence} to facilitate the visualization and comparison of the data between the different cases. \srebuttal{The Figures \ref{fig:ulaac_atb} and \ref{fig:ulaac_bta}} are analogous: Each column represents a different PV orientation, while the rows correspond to the flow split ratio under consideration. \srebuttal{In addition, both figures show the direction of the time-averaged WSS vector, which helps to interpret the stagnation regions near the wall.}

 $\text{RT}/\text{RT}_{\text{max}}$ values in \srebuttal{Figure \ref{fig:ulaac_atb}}, validate the trend observed in \rebuttal{Sections} \S \ref{subsec:LAA_RT} and \S \ref{stag_vol_evolution} for the AB transition. All values progressively decrease, reaching a minimum in case AB3. The highest RT values are concentrated in a narrow band located at the tip of the LAA. It should be mentioned that the $\text{RT}/\text{RT}_{\text{max}}$ values are higher for the 59-41\%LPV-RPV flow split, following the type A atrial behavior. This tendency is reversed by modifying the RPV orientation (shown in Cases AB3.1-3.3) and completing the AB transition.

TAWSS maps show the opposite trend compared to the $\text{RT}/\text{RT}_{\text{max}}$ maps, showing low TAWSS values where the $\text{RT}/\text{RT}_{\text{max}}$ showed high values. The lowest TAWSS values can be found in a narrow band near the LAA tip, whereas the highest values are located close to the ostium. Finally, OSI presents low values for all cases, indicating minimal flow oscillations barely affected by modifications in the PV orientations. 

Regarding the BA transition \srebuttal{(Figure \ref{fig:ulaac_bta})}, there is a progressive increase in $\text{RT}/\text{RT}_{\text{max}}$ values in the BA1-BA3 cases. These cases show the characteristic trend of type B, as evidenced by the higher $\text{RT}/\text{RT}_{\text{max}}$ values observed for the flow split ratio 37-63\%LPV-RPV. In contrast, the BA2.1-BA2.3 cases show a reverse flow split trend, indicating a complete transformation to type A. Between these BA2.1-BA2.3 cases, the lowest $\text{RT}/\text{RT}_{\text{max}}$ value is shown in the BA2.3 case. It should be noted that after the transition BA the patient presents higher $\text{RT}/\text{RT}_{\text{max}}$ values than in the base case. As for the previous transition, the TAWSS values show the opposite trend compared to $\text{RT}/\text{RT}_{\text{max}}$. The lowest values of TAWSS are found in a narrow band that corresponds to the LAA tip. Minimum values are reached in the flow split ratio 59-41\% LPV-RPV for cases BA2.1-BA2.3. Lastly, the oscillations in the flow direction are more affected by changes in the PV orientation and flow split ratio than in the previous transition. The lowest OSI values are shown in the base case (where the lowest $\text{RT}/\text{RT}_{\text{max}}$ values and the highest TAWSS values are found), while the highest OSI values are concentrated in a region close to the tip of the LAA for BA1-BA3 cases with a flow split ratio of 37-63\%LPV-RPV, and for BA2.1-BA2.3 cases with a flow split ratio of 59-41\%LPV-RPV.

\section{Discussion}
\label{sec:discussion}

Recent advances in computational modeling and medical imaging techniques and the growth in processing power make it possible to simulate patient-specific high-fidelity models, empower interdisciplinary collaborations, and drive a revolution in cardiovascular research. These synergies enable the generation of increasingly realistic digital twins to investigate pathological mechanisms, support the design and implantation of medical devices, and improve cardiac diagnosis. 

With respect to AF, CFD techniques have allowed in recent years to deepen our understanding of the hemodynamic substrate in patients with AF (\cite{duenas2024reduced, corti2022impact, garcia2021demonstration}). Together with patient-specific atrial segmentations, flow boundary conditions in the PV/MV, and wall motion reconstruction, these simulations provide an accurate 3D time-resolved representation of atrial flow. This time-resolved flow field allows calculation of hemodynamic indices related to blood stasis \cite{duenas2022morphing, gonzalo2022non,pons2022joint}, such as residence time (RT) or wall shear-derived indices (TAWSS, OSI, etc.). The analysis of these indices is especially relevant within the LAA, where thrombi are more prone to form. 

Most of the previous atrial CFD studies focused on the relationship between blood stasis and LAA morphology \cite{Garcia-Isla2018,Masci2019,garcia2021demonstration}, on contrasting different atrial modeling hypotheses \cite{duenas2021comprehensive, duran2023pulmonary,gonzalo2022non,Mill2021, Khalili2024}, or on looking for the most suitable metrics to quantify stasis \cite{duenas2021comprehensive, corti2022impact, Achille2014}. However, the translation of computational advances to improve clinical decisions about AF still faces significant challenges. For example, the formation of a stagnant region within the LAA is highly dependent on atrial flow patterns and, particularly, on the position of the main atrial vortex in relation to the LAA \cite{rodriguez2024influence}. However, atrial flow patterns depend on several parameters: LA/LAA morphology, PV orientations, cardiac output, flow split ratio, wall motion, etc. Therefore, understanding the influence of each of these parameters will require isolating and studying its influence one by one instead of performing patient-specific simulations in which several parameters are simultaneously modified. Furthermore, the variation of these parameters depends on time: atrial remodeling can alter the LA/LAA morphology over months/years \cite{boyle2021fibrosis}, while atrial wall motion and cardiac output can vary significantly throughout the day due to changes in cardiac conditions \cite{TAKAGI2012,FARESE2019,Elliott2023} (rest/exercise, sinus/fibrillated rhythm). Even the flow split ratio is highly dependent on the position of the patient \cite{duran2023pulmonary,WRAPU19}. For these reasons, although these studies have contributed to a better understanding of AF phenomena and the boundary conditions that best reproduce atrial flow, they have not yet gone far enough to build a predictive understanding of each of their anatomical and functional determinants.

To meet the challenges described above, this manuscript will focus on achieving a better understanding of the influence of the PV orientations on the formation of the main atrial vortex and its implications on stasis within the LAA. Although it is confirmed that PV orientations affect stasis within the LAA \cite{duenas2022morphing,Mill2021_pulmonary, mill2023role}, the number of CFD studies that include it as a parameter to assess stroke risk is reduced. Mill \textit{et al.} \cite{Mill2021_pulmonary, mill2023role} studied several geometries, reporting the importance of the PV angles on atrial flow patterns. However, their simulation results combined several patient-specific modifications (LAA and LA morphologies, PV orientations), so it was impossible to analyze the isolated effect of PV orientations. Dueñas-Pamplona \textit{et al.} \cite{duenas2022morphing} made a preliminary parametric approximation on a patient-specific basis to study the influence of PV on stagnant LAA volume. However, they only considered extreme PV configurations to quantify the maximum variation of stagnant volume without performing a gradual and exhaustive PV variation to understand its effect on different flow patterns. In fact, neither the formation of the main circulatory flow nor the various influences of varying LPV and RPV were studied. 

To fill this gap, this work combined some recent methodological advances to better understand the relationship between PV orientations and the formation of the main atrial vortex, together with its implications for LAA stasis. \rebuttal{The results highlight how the position of the main atrial vortex is affected by changes in the PV orientations and also, although to a lesser extent, by changes in the flow split ratio}.

As is well known, the LPV flow is characterized by the formation of the main atrial vortex, while the RPV flow is mostly conducted directly towards the MV \cite{Vedula2015}. Our departure point was LA segmentations from two patients with AF (Figure \ref{fig:cases_parts}, Table \ref{tab:clinic-data}). Although we know this is a limited number of patients, the sample contained two extreme representative atrial flow patterns that can be identified depending on the position of the main atrial vortex \cite{rodriguez2024influence}. Patient 1 presents a type A flow behavior, with the main atrial vortex forming just after entering the LA and near the LPV, while Patient 2 presents a type B flow behavior, with the main atrial vortex being pushed toward the LA roof and near the RPV. It should be noted that this flow classification also implies different responses to changes in the flow split ratio: type A atriums will present a better LAA washing for a flow split ratio of 37-63\%, while type B atriums will present a better LAA washing for a flow split ratio of 59-41\%. 

Second, a morphing methodology \cite{duenas2022morphing, duenas2024reduced} was applied to perform a geometric data augmentation. Specifically, the PV were rigged and skinned following the method proposed by Dueñas-Pamplona \textit{et al.} \cite{duenas2022morphing} to obtain different PV orientations. To demonstrate in a parametric manner how the PV orientations play a fundamental role in the formation of the atrial flow patterns, the LPV and RPV orientations were gradually modified to transition Patient 1 from Type A to Type B (transition AB) and Patient 2 from Type B to Type A (transition BA). Completing these transitions highlights how the different flow patterns governing LAA washout are highly affected by PV orientation, as well as \rebuttal{allows the separate study} of LPV and RPV.
To achieve these transitions, two transformations should be performed: (i) the main atrial vortex should be displaced toward its new position by changing the LPVs orientation, and (ii) the RPVs should be oriented toward the main circulatory flow (transition AB) or towards the MV (transition BA). A methodology similar to that proposed by Mill \textit{et al.} \cite{mill2023role} was used to parameterize these transformations and measure the PV characteristic angles. In this work, we selected coronal and axial planes as a reference to define the characteristic angles, shown in Figure \ref{fig:angles}. This choice allowed us to generalize the proposed PV parameterization, as these planes are widely used in clinical practice and will facilitate comparison between different anatomies. The complete methodology is explained in detail in \S \ref{subsec:morphological_descriptors}, and the PV transformations are detailed in Figures \ref{fig:geoms_atb} and \ref{fig:geoms_bta}, and Tables \ref{tab:angle-data1} and \ref{tab:angle-data2}. To isolate the influence of PV orientations, the rest of the parameters were kept constant during these transformations: cardiac output, atrial motion, LAA morphology, etc.

Depending on the patient-specific distribution of fibrosis, the motion of the atrial wall is altered during episodes of AF. To prevent this effect from introducing bias into our analysis and to homogenize the effect of fibrosis between patients, we used the same kinematic model as Rodríguez-Aparicio \textit{et al.} \cite{rodriguez2024influence}, which was first proposed by \cite{corti2022impact, zingaro2021hemodynamics,zingaro2021geometric}. This model allows \rebuttal{generating} a smooth wall motion according to the patient-specific transmitral flow profile. Although this model may not capture all patient-specific flow characteristics, its efficacy in replicating LA and LAA hemodynamics is demonstrated \cite{kjeldsberg2024impact}, facilitating the isolation of geometrical and functional parameters and thus the study of PV orientations. 

Regarding the rheology of the blood, we considered constitutive relations based on the Carreau model, which was widely used in previous CFD studies of cardiovascular flows \cite{gonzalo2022non, Albors2023rheological, Agujetas2018}. Although Newtonian fluid assumption is widely used for cardiac cavities, as non-Newtonian atrial effects are reduced \cite{Garcia-Isla2018,Mill2020}, recently Gonzalo \textit{et al.} \cite{gonzalo2022non} suggested that Newtonian effects can affect hemodynamics within the LAA. These effects would shape the filling and draining jets, affecting the secondary swirling motions that govern LAA washing when LA wall motion is impaired.

The gradual flow transitions AB and BA were monitored by analyzing the transformations achieved in the blood streamlines, shown in Figure \ref{fig:velocity_streamlines} for 90\%RR, when the atrial flow structures are fully developed. Although it was already known that the LPV flow governs the main atrial vortex \cite{vedula2016effect, rodriguez2024influence}, to the best of our knowledge, this is the first time that it has been demonstrated how the position of its center can be modified by changing the orientations of the LPVs. Moreover, the influence of RPVs is minor but not negligible. Their orientation determines the nature of \rebuttal{the interaction between the RPV flow and} the main atrial vortex: \rebuttal{it} can either surround it and flow directly towards the MV or collide against it, generating a much more chaotic flow pattern. This chaotic flow pattern was previously observed \cite{rodriguez2024influence}, but the magnitude of this interaction was not previously studied by isolating the effect of gradually varying the PV orientations. 

The transformation of the flow patterns was then confirmed by analyzing the vortex structures, whose evolution for both transitions is shown in Figure \ref{fig:vor_vortex_evo}. We calculated the criterion $q_2$, showing the vortex structures for $q_2 = 0.2$. This criterion is widely used to visualize vortex structures and assess flow characteristics \cite{Hunt1988,Jeong1995}. It was also used before for atrial flow \cite{chnafa2014image, Otani2016, rodriguez2024influence}. Some representative time instants were selected to show the temporal evolution of the atrial flow structures for both transitions. The result confirmed that the maximum vorticity values are reached near the PV \cite{Otani2016}. The main vortex structures are fully developed around 80\%RR, and their location depends on the type of atrial flow. It can also be observed how the collision of the RPV flow with the main flow structure leads to a disordered and chaotic vortical structure throughout the atrial domain. These observations ratify previously observed flow patterns \cite{rodriguez2024influence}, confirming at the same time that the transitions AB and BA were completed.

\srebuttal{Finally, the effect of variations in the PV orientations on LAA washing was assessed by means of hemodynamic indices. Both volumetric and surface indices were assessed.  The volumetric RT data provide a complete overview of the stagnant state of a particular LAA. However, since it is a 3D scalar field, it is difficult to interpret its meaning directly. In addition, the complexity and variability of atrial morphology make it difficult to compare this 3D RT field between different patients and cardiac conditions. For these reasons, the stagnant volume was calculated to translate the information from the 3D scalar field into a single scalar, making it easier to plot and compare the different cases (Figure \ref{fig:stagnant_volume_evolution} and Tables \ref{tab:stagnant_evolution_atb} and \ref{tab:stagnant_evolution_bta}). On the other hand, the recent theory of WSS Lagrangian Coherent Structures (WSS LCS) has suggested that near-wall mass transport is mainly dictated by the WSS vector \cite{arzani2018wall, arzani2016lagrangian}, that is, by the near-wall velocity field. This near-wall mass transport is particularly relevant in the context of thrombosis with thin concentration boundary layers for thrombin. To consider this effect, the wall results were projected onto the ULAAC system, facilitating the comparison of the near-wall RT with WSS-derived indices such as TAWSS and OSI. The direction of the time-averaged WSS vector was displayed along with the TAWSS contours, helping to interpret the near-wall stagnant regions \cite{arzani2018wall} (Figures \ref{fig:ulaac_atb} and \ref{fig:ulaac_bta}). Although both volumetric and surface indices ultimately depend on the volumetric velocity field, we believe that both approaches are interesting and show complementary information for the study of thrombogenesis.}

The gradual AB and BA transitions produced important changes in LAA stasis. Figure \ref{fig:LAA_RT_by_transformation} shows clear patterns for $\text{RT}/\text{RT}_{\text{max}}$. The region with higher $\text{RT}/\text{RT}_{\text{max}}$ values in the AB transition (Figure \ref{subfig:rt_atb}) is affected by the position of the main atrial vortex (cases AB0-AB5), showing a minimum for the AB3 case. Regarding the influence of the flow split ratio, is noticeable but less important than that of the vortex position. The RT pattern is still type A, so RT is reduced when changing the flow split ratio from 59-41\% to 37-63\%. On the other hand, once RPV orientation is modified, the $\text{RT}/\text{RT}_{\text{max}}$ values are affected by the magnitude of the interaction between the RPV flow and the main atrial vortex (cases AB3.1-AB3.3). It can be seen that as the RPV flow is gradually oriented towards the position of the main atrial vortex, this flow interaction worsens the LAA washing and increases $\text{RT}/\text{RT}_{\text{max}}$. When the AB transition is complete, the LAA washing is governed by RPV flow for a flow split ratio of 59-41\%, becoming more chaotic and worsening the LAA washing for 37-63\%. An analogous RT pattern can be seen for the BA transition (Figure \ref{subfig:rt_atb}).  The $\text{RT}/\text{RT}_{\text{max}}$ is affected by the position of the main atrial vortex (cases BA0-BA3), showing a minimum for the base case. The transition BA is completed when the RPV are oriented to the MV, allowing the development of the main circulatory flow towards the ostium and improving the LAA washing when changing the flow split ratio from 59-41\% to 37-63\%. 

\rebuttal{As mentioned}, the stagnant blood volume was calculated for each different case (Figure \ref{fig:stagnant_volume_evolution} \rebuttal{and Tables \ref{tab:stagnant_evolution_atb} and \ref{tab:stagnant_evolution_bta}}), showing a clear pattern for both transformations, analogous to the one commented before for $\text{RT}/\text{RT}_{\text{max}}$. Both transitions present smooth variations in the stagnant volume when varying the position of the main atrial vortex, with significant changes when the RPV orientation is modified and the transitions are completed. For better visualization and comparison, the wall results of $\text{RT}/\text{RT}_{\text{max}}$, TAWSS, and OSI were projected onto the ULAAC system (\srebuttal{Figures \ref{fig:ulaac_atb} and \ref{fig:ulaac_bta}}). The patterns displayed for $\text{RT}/\text{RR}$ and TAWSS are complementary: \rebuttal{Higher values of $\text{RT}/\text{RR}$ are concentrated in a narrow band near the LAA tip, which coincides with lower values of TAWSS. This relationship between the high RT and low TAWSS regions was expected, as blood tends to stagnate in regions with low velocities \cite{duenas2024reduced, rodriguez2024influence}.} Changes in the direction of flow within the LAA are more notable for the BA transition, although the OSI remains at low values. The cases with higher $\text{RT}/\text{RT}_{\text{max}}$ have higher OSI values concentrated near the tip of the LAA, which is consistent with previous work \cite{chen2023hemodynamic}. This effect may be due to the combination of larger curvatures and low RT values present in this region.

Our work confirmed that the proximity of the main atrial vortex to the ostium can induce secondary swirling flows within the LAA that facilitate washing and prevent blood stagnation, which is consistent with previous work \cite{duran2023pulmonary, gonzalo2022non}. However, the position of the main atrial vortex can be significantly modified by changes in the PV orientations and, to a lesser extent, by the flow-split ratio. We also confirmed that different types of atrial flow patterns have different LAA washing behaviors and responses to changes in the flow split ratio \cite{rodriguez2024influence}. For the first time, we identified the relationship between the main atrial vortex and its interaction with the RPV flow, showing how this can be altered by changing the PV orientations, which, to the best of our knowledge, has not been done before. Moreover, we demonstrated how the atrial flow pattern and its response to changes in the flow split ratio can be totally reverted by an appropriate combination of transformations in the RPV and LPV orientations. 

\subsection{Clinical applications}
\label{subsec:clinical}

In recent years, CFD techniques have shown their value in shedding light on multiple cardiovascular diseases, and even helping during cardiac diagnosis \cite{Rigatelli2022, Morris2016}. Regarding AF, CFD contributed to quantifying stasis within LAA in a patient-specific basis and achieving a better understanding of atrial flow patterns and the prothrombotic substrate of the disease \cite{Garcia-Isla2018, garcia2021demonstration, duenas2021comprehensive, corti2022impact}. This may facilitate the development of new tools to improve both diagnosis and treatment. However, while CFD studies have improved our understanding of AF mechanisms, atrial flow, and the determination of otherwise inaccessible stasis-related parameters, the connection between atrial flow patterns and stasis in LAA remains unclear. This study aimed to investigate atrial flow patterns and their relationship to LAA washing, and more specifically how these were affected by variations in the PV orientations.

The findings of this study could have direct implications for the diagnosis of AF, as PV orientations are not taken into account during standard clinical practice when assessing the risk of stroke in a particular patient. We suggest introducing PV orientations in the characterization of the patient-specific risk of stroke, as it can affect the risk of thrombosis in the medium and long term. Moreover, we have confirmed that the fact that a patient sleeps recurrently on her/his left/right side affects the flow split ratio and thus may modify the atrial flow patterns and the risk of stasis \cite{rodriguez2024influence}.

\subsection{\rebuttal{Limitations and future work}}
\label{subsec:limitations}

\rebuttal{The use of a kinematic model for AF simulation may limit the patient-specificity of the models, which may not reflect the full patient-specific characteristics. However, the purpose of the study was not to achieve full patient specificity but to better understand the fundamental role of PV orientation in the formation of atrial flow patterns. Our aim was to systematically quantify the effect of PV orientations on blood stasis within the LAA. With this in mind, using a kinematic model allowed us to isolate the effect of local fibrotic patterns from the anatomical effects, facilitating the comparison between different morphed geometries. This approach offers a useful modeling alternative \cite{zingaro2021geometric,zingaro2021hemodynamics, corti2022impact}, and its efficacy in replicating the LA and LAA hemodynamics was recently demonstrated \cite{kjeldsberg2024impact}. For all these reasons, we considered it appropriate to isolate geometric and functional parameters to study the effect of PV orientations on blood stasis in the LAA.}

\rebuttal{Clinical images of two representative patients with AF were used for the study. Although the small number of the cohort did not allow us to study the influence of factors such as the influence of the LAA morphological variations or the AF type, the morphing applied to modify the PV orientations allowed us to increase the number of cases to 18 geometries, which means a total number of 36 simulations (2 flow split ratios per geometry). We believe that this data-augmented cohort was large enough to meet the objective of the study: to analyze the influence of PV orientations on LAA washing, isolating the rest of the morphological and flow-related parameters. Future work will focus on increasing the existing cohort to a higher number of patients to confirm and generalize the results achieved so far.}

\rebuttal{Along this line, an important step towards taking PV orientations into account in routine clinical practice for patients with AF would be developing a predictive model. This model should be able to predict the position of the main atrial vortex and stasis inside the LAA based on the PV orientation for a particular patient. The construction of such a model would involve working with a large patient base to perform a data augmentation procedure with the methodology described here.}

Similarly to most previous numerical studies of LA, an MV model was not considered \cite{Otani2016, garcia2021demonstration, duenas2021comprehensive, bucelli2022mathematical}. This assumption is justified by whole-left-heart simulations, which suggest that the dynamics of MV does not affect atrial flow patterns \cite{Vedula2015}.

Biochemical modeling of thrombus formation was not included in our blood analysis. It would have required considering multiple biochemical equations \cite{seo2016coupled,RMN18}, increasing the computational burden. In addition, their precision is still limited due to the \rebuttal{difficulty of fitting these equations to specific patient conditions \cite{rodriguez2024influence}.}

\subsection{Conclusion} 
\label{subsec:conclusion}

\rebuttal{The goal of this research was to improve understanding of how PV orientations interact with atrial flow patterns and influence blood stasis in the LAA.} To this end, some of the most recent techniques in the field were applied: an atrial morphing process to analyze the isolated effect of modifying the PV orientations, an atrial kinematic model to isolate anatomical from functional effects, and projection of the wall results on the ULAAC system to facilitate visualization and comparison of data between different \rebuttal{atrial morphologies}. \rebuttal{The findings demonstrate systematically how alterations in LPV orientations can significantly alter the position of the main atrial vortex, which to the best of our knowledge has not been done before. Moreover, the results highlight the important role of RPV orientation in the interaction between the RPV flow and the main atrial vortex, crucial to define the LA patterns and, thus, the LAA washing. Moreover, we demonstrated for the first time how the flow patterns response to changes in the flow split ratio can be totally changed by a combination of controlled modifications in the PV orientations.}

\section*{\color{black}CRediT authorship contribution statement}
\srebuttal{\textbf{Sergio Rodríguez-Aparicio}: Investigation, Validation, Data Curation, Visualization, Writing - Original Draft. \textbf{Conrado Ferrera}: Writing - Original Draft, Writing - Review \& Editing, Supervision. \textbf{María Eugenia Fuentes-Cañamero}: Resources. \textbf{Javier
García García}: Writing - Review \& Editing. \textbf{Jorge Dueñas-Pamplona}: Conceptualization, Methodology, Software, Visualization, Writing - Original Draft, Writing - Review \& Editing, Supervision.}

\section*{Acknowledgments}

This work was supported by Junta de Extremadura and
FEDER funds under project IB20105. We thank the Programa Propio - Universidad Polit\'{e}cnica de Madrid, and the Ayuda Primeros Proyectos de Investigaci\'{o}n ETSII-UPM. We also thank Programa de Excelencia para el Profesorado Universitario de la Comunidad de Madrid for its financial support and the CeSViMa UPM project for its computational resources.









\appendix
\section{\rebuttal{Supplementary Data}}
\rebuttal{The following is the Supplementary Data related to this article.}
\begin{itemize}
    \item \rebuttal{A.1: Atrial morphologies before and after the morphing procedure. It consists of the following figures:} 
    \begin{itemize}
        \item \rebuttal{\textbf{Figure A.1.1}: Atrial morphologies for transition AB.}
        \item \rebuttal{\textbf{Figure A.1.2}: Atrial morphologies for transition BA.}
    \end{itemize}
    \item \rebuttal{A.2 Blood flow streamlines at 90\%RR, showing the position of the main circulatory flow. It consists of the following animations:}
        \begin{itemize}
        \item \rebuttal{\textbf{Animation A.2.1}: Blood flow streamlines for transition AB.}
        \item \rebuttal{\textbf{Animation A.2.2}: Blood flow streamlines for transition BA.}
    \end{itemize}
    \item \rebuttal{A.3 Vortex structure evolution during systole. It consists of the following animations:}
            \begin{itemize}
        \item \rebuttal{\textbf{Animation A.3.1}: Vortex structure evolution for transition AB.}
        \item \rebuttal{\textbf{Animation A.3.2}: Vortex structure evolution for transition BA.}
    \end{itemize}
\end{itemize}

\printcredits


\bibliographystyle{unsrt}

\bibliography{cas-dc.bbl}

\bio{}
\endbio

\endbio

\end{document}